\documentclass[aps,rmp,square]{revtex4}
\usepackage{amsmath,amssymb}
\usepackage{graphicx}
\usepackage{pst-plot}
\usepackage{mathrsfs}
\usepackage{enumerate}

\usepackage{pstricks}
\usepackage{pst-func}
\usepackage{refcount}
\usepackage{cases}

\allowdisplaybreaks

\newtheorem{my-theorem}{Theorem}

\def\ocite#1{[\citenum{#1}]}

\setcitestyle{numbers}

\begin{document}

\title{Refutation of Some Arguments Against my Disproof of Bell's Theorem}

\author{Joy Christian}

\email{jjc@alum.bu.edu}

\affiliation{Wolfson College, University of Oxford, Oxford OX2 6UD, United Kingdom}

\begin{abstract}
In a couple of recent preprints Moldoveanu has suggested that there are errors in my disproof
of Bell's theorem. Here I show that this claim is false. In particular, I show that my
local-realistic framework is incorrectly and misleadingly presented in both of his preprints.
In addition there are a number of serious mathematical and conceptual errors in
his discussion of my framework. For example, contrary to his claim, my
framework is manifestly non-contextual. In particular, quantum correlations are understood within it as
purely topological effects, not contextual effects.
\end{abstract}

\maketitle

\parskip 5pt

\baselineskip 3pt

\section{Introduction}

Bell's theorem \ocite{Bell-1964-666} is based on an assumption that in any correlation experiment local
measurement results can be described by functions of the form
\begin{equation}
\pm\,1\,=\,
{\mathscr A}({\bf n},\,\lambda): {\rm I\!R}^3\!\times\Lambda\longrightarrow {\cal I}\subseteq{\rm I\!R}\,, \label{non-map-999}
\end{equation}
with ${{\rm I\!R}^3\ni {\bf n}}$ representing a space of experimental contexts, ${\lambda\in \Lambda}$ representing
a complete initial state of the system, and ${{\cal I}\subseteq{\rm I\!R}}$ representing the set of all possible measurement
results in question \ocite{Bell-1964-666}. Elsewhere I have shown that this assumption is false
\ocite{disproof}\ocite{Christian-666}\ocite{Restoring}\ocite{What-666}\ocite{illusion-666}\ocite{photon-666}\ocite{reply}\ocite{Further-666}\ocite{experiment-666}. Elementary topological scrutiny reveals that no such
local function---or its
probabilistic counterpart ${P({\mathscr A}\,|\,{\bf n},\,{\lambda})}$---is capable of providing a complete
account of every possible measurement result, even for the simplest of quantum systems.
Unless enumerated by functions of the topologically correct form
\begin{equation}
\pm\,1\,=\,{\mathscr A}
({\bf n},\,\lambda): {\rm I\!R}^3\!\times\Lambda\longrightarrow S^3 \hookrightarrow{\rm I\!R}^4, \label{fon-map-6666}
\end{equation}
with their codomain ${S^3}$ being a parallelized 3-sphere, it is not possible to account for every possible measurement result
for any two-level quantum system \ocite{What-666}.
More specifically, unless the measurement results of Alice and Bob are represented by the equatorial points of a
parallelized 3-sphere, the completeness criterion of EPR is not satisfied, and then there is no meaningful Bell's theorem to begin
with \ocite{Restoring}\ocite{What-666}\ocite{illusion-666}\ocite{photon-666}\ocite{reply}\ocite{experiment-666}.
In fact, na\"ively replacing the
{\it simply-connected} codomain ${S^3}$ in the above function by a {\it totally-disconnected} ${\,}$set
${S^0\equiv\{-1,\,+1\}}$, as routinely done within all Bell type arguments, is a guaranteed way of introducing incompleteness
in the accounting of measurement results from the start \ocite{What-666}\ocite{illusion-666}. As I have argued elsewhere
\ocite{disproof}\ocite{Christian-666}\ocite{Restoring}\ocite{What-666}\ocite{illusion-666}\ocite{photon-666}\ocite{reply}\ocite{Further-666}\ocite{experiment-666},
the {\it only} ${\,}$unambiguously complete way of\break accounting for every possible measurement result locally
is by means of standardized variables or bivectors of the form
\begin{equation}
{\rm I\!R}^4\hookleftarrow S^3 \supset S^2 \ni {\boldsymbol\mu}\cdot{\bf n}\,\equiv\,\lambda\,I\cdot{\bf n}\,
=\,\pm\,1\;\,{\rm about}\,\,{\bf n}\in{\rm I\!R}^3
\subset{\rm I\!R}^4,
\end{equation}
for such bivectors intrinsically represent the equatorial points of a parallelized 3-sphere. 
Here ${{\boldsymbol\mu}=\lambda\,I}$ is a hidden variable or the complete state of the system,
with ${\lambda\equiv\pm\,1}$ being a fair coin
and ${I={{\bf e}_x}{{\bf e}_y}{{\bf e}_z}}$ being the fundamental volume form of the physical space. In statistical
terms these bivectors then represent the standard scores, which are mathematical counterparts of the actually
observed raw scores: ${+1}$ or ${-1}$ \ocite{disproof}\ocite{Restoring}. Moreover, once
parallelized by a field of such bivectors (and their extensions to ${{\rm I\!R}^4}$), a 3-sphere remains as closed under
multiplication of its points as the 0-sphere: ${\{-1,\,+1\}}$. As a result, setting the codomain of the standard scores
to be the space of bivectors---which is isomorphic to an  equatorial 2-sphere within a parallelized 3-sphere---guarantees
that the locality or factorizability condition of Bell is automatically satisfied, for any number of measurement settings.
The resulting model \ocite{disproof}\ocite{Restoring}
of the EPR-Bohm correlations is therefore complete, local, and realistic, in the
precise senses defined by EPR and Bell \ocite{illusion-666}.

The informal picture that emerges from these formal arguments is as follows: EPR-Bohm correlations are
telling us that we live in a parallelized 3-sphere, which differs from our usual conception of the physical space as
${{\rm I\!R}^3}$ only by a single point at infinity \ocite{Restoring}\ocite{What-666}. All measurement results, such as
${{\mathscr A}({\bf a},\,{\lambda})=\pm\,1}$, ${{\mathscr B}({\bf b},\,{\lambda})=\pm\,1}$, etc., are simply
detections of one of the two possible orientations---or one of the two possible senses of local rotations---of this 3-sphere,
predetermined by the initial state ${\lambda=\pm\,1}$. In other words, the hidden variable in this picture is the initial
orientation of the physical space itself, which predetermines all possible outcomes at all possible measurement directions in
the EPR-Bohm scenario. As a result, the measurement results are not contextual in any sense. For example, for a given orientation
${\lambda}$ of the 3-sphere, the actual value ${\mathscr A}$
observed at a given direction ${\bf a}$, whether it is ${+1}$ or ${-1}$, does
not change at all if, say, the measurement setting is changed from ${{\bf a}}$ to ${{\bf a}'}$. Why do the correlations between
such purely random, non-contextual outcomes then turn out to be sinusoidal rather than linear? The reason has to do with the fact
that there is a
non-trivial twist in the Hopf fibration of the 3-sphere \ocite{Restoring}\ocite{experiment-666},
and this twist is responsible for making the correlations
more disciplined (or stronger) than linear. Consequently, in this picture the EPR-Bohm correlations are ontologically no different
from those between the changing colors of Dr.${\;}$Bertlmann's socks discussed by Bell \ocite{Sock}.

As elegant, compelling, and conceptually appealing this picture is, in a pair of recent preprints Moldoveanu has questioned its
validity by suggesting that there are errors in my disproof of Bell's theorem \ocite{M-1}\ocite{M-2}. In what follows I
show that this claim is false. The errors are in fact in Moldoveanu's understanding of my local-realistic framework. Contrary to
what is claimed in his preprints, my work on the subject,
\ocite{disproof}\ocite{Christian-666}\ocite{Restoring}\ocite{What-666}\ocite{illusion-666}\ocite{photon-666}\ocite{reply}\ocite{Further-666}\ocite{experiment-666}, is perfectly cogent and error free. In what follows
I bring out a number of mathematical and conceptual errors in his discussion of my work.

\section{Orientation of the 3-Sphere is Not a Convention But a Hidden Variable}

The first issue raised by Moldoveanu has to do with the hidden variable duality relation I have used in most of my
papers as a convenient calculational tool:
\begin{equation}
{\bf a}\wedge{\bf b}={\boldsymbol\mu}\,\cdot({\bf a}\times{\bf b})\equiv\lambda\,I\cdot\,({\bf a}\times{\bf b})\,.\label{1-on-1}
\end{equation}
Here ${{\boldsymbol\mu}=\lambda\,I}$ is a hidden variable of the model, with ${\lambda\equiv\pm\,1}$ as a fair coin
and ${I={{\bf e}_x}{{\bf e}_y}{{\bf e}_z}}$ the standard volume form
of the physical space. It represents the two alternately possible orientations of the 3-sphere, or equivalently the
two alternately possible handedness of the bivector subalgebra representing the points of the 3-sphere. If we separate out the
two independent instances of ${\lambda}$, then the combined duality relation (\ref{1-on-1}) splits into two independent relations,
with ${{\bf a}\wedge{\bf b}=+\,I\cdot\,({\bf a}\times{\bf b})}$ representing the duality relation for the right-handed bivector
basis and ${{\bf a}\wedge{\bf b}=-\,I\cdot\,({\bf a}\times{\bf b})}$ representing the duality relation for the left-handed bivector
basis. Although this differs from the usual practice in geometric algebra, the combined duality relation provides considerable
computational ease in my model, provided one does not misunderstand its correct meaning. It leads to the following identity,
which is a very convenient and powerful mathematical tool with
rich physical and geometrical meaning, and plays a central role in most of my papers:
\begin{equation}
(\,{\boldsymbol\mu}\cdot{\bf a})(\,{\boldsymbol\mu}\cdot{\bf b})\,
=\,-\,{\bf a}\cdot{\bf b}\,-\,{\boldsymbol\mu}\cdot({\bf a}\times{\bf b}),\label{bi-identityyyyy}
\end{equation}
Since the numbers ${(\,{\boldsymbol\mu}\cdot{\bf a})}$ and ${(\,{\boldsymbol\mu}\cdot{\bf b})}$ are identified in the model
as the statistically pertinent standard scores (as opposed to actually observed raw scores, ${+1}$, ${-1}$, etc.),
the EPR correlation follows at once form the above identity:
\begin{align}
{\cal E}({\bf a},\,{\bf b})\,=\lim_{\,n\,\gg\,1}\left[\frac{1}{n}\sum_{i\,=\,1}^{n}\,
(\,{\boldsymbol\mu}^i\cdot{\bf a})(\,{\boldsymbol\mu}^i\cdot{\bf b})\right]
\,=\,-\,{\bf a}\cdot{\bf b}\,-\lim_{\,n\,\gg\,1}\left[\frac{1}{n}\sum_{i\,=\,1}^{n}\,
{\lambda}^i\,I\cdot({\bf a}\times{\bf b})\right]\,=\,-\,{\bf a}\cdot{\bf b}\,+\,0\,.
\end{align}
Note also that since ${({\bf a}\times{\bf b})}$ is an exclusive direction yielding a null result, the last summation is in fact
unnecessary.

Rather surprisingly, this convenient mathematical tool has managed to confuse many people over the years for reasons
that are only partly understandable. To be sure, in the standard practice the duality relation such as
${{\bf a}\wedge{\bf b}=+\,I\cdot\,({\bf a}\times{\bf b})}$ is a fixed convention, and holds true whether one uses right-handed
vector basis or left-handed. It simply tells us how the plane of ${{\bf a} \wedge {\bf b}}$ is related to its orthogonal vector
${{\bf a} \times {\bf b}}$. It should be remembered, however, that, although based on geometric algebra, mine is primarily a hidden
variable model. Moreover, it should be remembered that the essence of my argument depends on the double cover property of the
physical space \ocite{experiment-666}, and therefore\break I am working primarily within the bivector subalgebra of the Clifford
algebra ${{Cl}_{3,0}}$ \ocite{disproof}\ocite{Restoring}. To make this clear, let me derive the indefinite duality
relation (\ref{1-on-1}) once again from the first principles, and show how it fits into the identity (\ref{bi-identityyyyy}).

Consider a right-handed frame of ordered basis bivectors,
${\{{\boldsymbol\beta}_x,\,{\boldsymbol\beta}_y,\,{\boldsymbol\beta}_z\}}$, and the corresponding bivector subalgebra
\begin{equation}
{\boldsymbol\beta}_j\,{\boldsymbol\beta}_k \,=\,-\,\delta_{jk}\,-\,\epsilon_{jkl}\,{\boldsymbol\beta}_l\label{lisr}
\end{equation}
of the Clifford algebra ${{Cl}_{3,0}}$. The latter is a linear vector space ${{\rm I\!R}^8}$ spanned by the orthonormal basis
\begin{equation}
\left\{1,\,\;{\bf e}_x,\,{\bf e}_y,\,{\bf e}_z,\,\;{\bf e}_x\wedge{\bf e}_y,\,
{\bf e}_y\wedge{\bf e}_z,\,{\bf e}_z\wedge{\bf e}_x,\,\;
{\bf e}_x\wedge{\bf e}_y\wedge{\bf e}_z\right\}\!, \label{cl30}
\end{equation}
where ${\delta_{jk}}$ is the Kronecker delta, ${\epsilon_{jkl}}$ is the Levi-Civita symbol, the indices
${j,\,k,\,l=x,\,y,}$ or ${z}$ are cyclic indices, and
\begin{equation}
{\boldsymbol\beta}_j\,=\,{\bf e}_k\wedge{\bf e}_l\,=\,I\cdot{\bf e}_j\,.\label{tog-13}
\end{equation}
Eq.${\,}$(\ref{lisr}) is a standard expression of bivector subalgebra,
routinely used in geometric algebra to define the right-handed frame of basis bivectors \ocite{Clifford}.
From (\ref{lisr}) it is easy to verify the familiar properties of the basis bivectors, such as
\begin{align}
({\boldsymbol\beta}_x)^2\,=\,({\boldsymbol\beta}_y)^2&\,=\,({\boldsymbol\beta}_z)^2\,=\,-\,1 \label{prop13} \\
\text{and}\;\;\;{\boldsymbol\beta}_x\,{\boldsymbol\beta}_y\,=
&\,-\,{\boldsymbol\beta}_y\,{\boldsymbol\beta}_x\;\;\;\text{etc.} \label{prop14}
\end{align}
Moreover, it is easy to verify that the bivectors satisfying the subalgebra (\ref{lisr})
form a right-handed frame of basis bivectors. To this end, right-multiply both sides of Eq.${\,}$(\ref{lisr}) by
${{\boldsymbol\beta}_l}$, and then use the fact that ${({\boldsymbol\beta}_l)^2=-1}$ to arrive${\;}$at
\begin{equation}
{\boldsymbol\beta}_j\,{\boldsymbol\beta}_k\,{\boldsymbol\beta}_l\,=\,+\,1\,. \label{rightH}
\end{equation}
The fact that this ordered product yields a positive value confirms that
${\{{\boldsymbol\beta}_x,\,{\boldsymbol\beta}_y,\,{\boldsymbol\beta}_z\}}$ indeed forms a right-handed frame of basis bivectors.
This is a universally accepted convention, found in any textbook on geometric algebra \ocite{Clifford}.

Suppose now ${{\bf a}=a_{\! j}\,{\bf e}_j}$ and ${{\bf b}=b_{\! k}\,{\bf e}_k}$ are two unit vectors in ${{\rm I\!R}^3}$,
where the repeated indices are summed over ${x,\,y,}$ and ${z}$.
Then the right-handed basis equation (\ref{lisr}) leads to
\begin{equation}
\{\,a_j\;{\boldsymbol\beta}_j\,\}\,\{\,b_k\;{\boldsymbol\beta}_k\,\}\,=\,-\,a_j\,b_k\,\delta_{jk}\,
-\,\epsilon_{jkl}\;a_j\,b_k\;{\boldsymbol\beta}_l\,,
\end{equation}
which, together with (\ref{tog-13}), is equivalent to 
\begin{equation}
(\,I\cdot{\bf a})(\,I\cdot{\bf b})\,
=\,-\,{\bf a}\cdot{\bf b}\,-\,I\cdot({\bf a}\times{\bf b}), \label{one-ident}
\end{equation}
where ${I={{\bf e}_x}{{\bf e}_y}{{\bf e}_z}}$ is the standard trivector. Geometrically this identity describes all points of a
parallelized 3-sphere.

Let us now consider a left-handed frame of ordered basis bivectors, which we also denote by
${\{{\boldsymbol\beta}_x,\,{\boldsymbol\beta}_y,\,{\boldsymbol\beta}_z\}}$. It is important to recognize,
however, that there is no {\it a priori} way of knowing that this new basis frame is in fact left-handed.
To ensure that it is indeed left-handed we must first make sure that it is an ordered frame by requiring
that its basis elements satisfy the bivector properties delineated in Eqs.${\,}$(\ref{prop13}) and (\ref{prop14}).
Next, to distinguish this frame from the right-handed frame defined by equation (\ref{rightH}), we must require
that its basis elements satisfy the property
\begin{equation}
{\boldsymbol\beta}_j\,{\boldsymbol\beta}_k\,{\boldsymbol\beta}_l\,=\,-\,1\,. \label{leftH}
\end{equation}
One way to ensure this is to multiply every non-scalar element in (\ref{cl30}) by a minus sign. Then,
instead of (\ref{tog-13}), we have
\begin{equation}
{\boldsymbol\beta}_j\,=\,-\,{\bf e}_k\wedge{\bf e}_l\,=\,(\,-\,I\,)\cdot(\,-\,{\bf e}_j\,)\,=\,I\cdot{\bf e}_j\,,\label{tog-20}
\end{equation}
and the condition (\ref{leftH}) is automatically satisfied.
As is well known, this was the condition imposed by Hamilton on his unit quaternions, which we now know are
nothing but a left-handed set
of basis bivectors \ocite{Clifford}. Indeed, it can be easily checked that the basis bivectors satisfying the properties
(\ref{prop13}), (\ref{prop14}), (\ref{leftH}), and (\ref{tog-20}) compose the subalgebra
\begin{equation}
{\boldsymbol\beta}_j\,{\boldsymbol\beta}_k \,=\,-\,\delta_{jk}\,+\,\epsilon_{jkl}\,{\boldsymbol\beta}_l\,. \label{lisr-left}
\end{equation}
Conversely, it is easy to check that the basis bivectors defined by this subalgebra do indeed form a left-handed frame. To this
end, right-multiply both sides of Eq.${\,}$(\ref{lisr-left}) by ${{\boldsymbol\beta}_l}$, and then use the property
${({\boldsymbol\beta}_l)^2=-1}$ to verify Eq.${\,}$(\ref{leftH}).
As is well known, this subalgebra is generated by the unit quaternions originally proposed by Hamilton \ocite{Clifford}. 
It is routinely used in the textbook treatments of angular momenta, but without mentioning the fact that it defines
nothing but a left-handed set of basis bivectors. It may look more familiar if we temporarily change notation and rewrite
Eq.${\,}$(\ref{lisr-left})${\;}$as
\begin{equation}
{\bf J}_j\,{\bf J}_k \,=\,-\,\delta_{jk}\,+\,\epsilon_{jkl}\,{\bf J}_l\,. \label{lisr-left-J}
\end{equation}
More importantly (and especially since Moldoveanu seems to have missed this point), I stress once again that there is no
way to set apart the left-handed frame of basis bivectors from the right-handed frame without appealing to the
intrinsically defined distinguishing conditions (\ref{rightH}) and (\ref{leftH}), or equivalently to the corresponding
subalgebras (\ref{lisr}) and (\ref{lisr-left}). Note also that at no time within my framework the two subalgebras
(\ref{lisr}) and (\ref{lisr-left}) are mixed in any way, either physically or mathematically.
They merely play the role of two distinct and alternative hidden variable possibilities. 

Suppose now ${{\bf a}=a_{\! j}\,{\bf e}_j}$ and ${{\bf b}=b_{\! k}\,{\bf e}_k}$ are two unit vectors in ${{\rm I\!R}^3}$, where
the repeated indices are summed over ${x,\,y,}$ and ${z}$. Then the left-handed basis equation (\ref{lisr-left}) leads to
\begin{equation}
\{\,a_j\;{\boldsymbol\beta}_j\,\}\,\{\,b_k\;{\boldsymbol\beta}_k\,\}\,=\,-\,a_j\,b_k\,\delta_{jk}\,
+\,\epsilon_{jkl}\;a_j\,b_k\;{\boldsymbol\beta}_l
\end{equation}
which, together with (\ref{tog-20}), is equivalent to 
\begin{equation}
(\,I\cdot{\bf a})(\,I\cdot{\bf b})\,
=\,-\,{\bf a}\cdot{\bf b}\,+\,I\cdot({\bf a}\times{\bf b}), \label{two-ident}
\end{equation}
where ${I}$ is the standard trivector. Once again, geometrically this identity describes all points of a
parallelized 3-sphere.

It is important to note, however, that
there is a sign difference in the second term on the RHS of the identities (\ref{one-ident}) and (\ref{two-ident}).
The algebraic meaning of this sign difference is of course clear from the above discussion, and it has been discussed
extensively in most of my papers, with citations to prior literature \ocite{Restoring}\ocite{What-666}\ocite{photon-666}.
But from the perspective of my model a more important question is: What does this sign difference mean {\it geometrically}${\,}$?
To bring out its geometric meaning, let us rewrite the identities (\ref{one-ident}) and (\ref{two-ident}) as
\begin{equation}
(\,+\,I\cdot{\bf a})(\,+\,I\cdot{\bf b})\,
=\,-\,{\bf a}\cdot{\bf b}\,-\,(\,+\,I\,)\cdot({\bf a}\times{\bf b}) \label{id-1}
\end{equation}
and
\begin{equation}
(\,-\,I\cdot{\bf a})(\,-\,I\cdot{\bf b})\,
=\,-\,{\bf a}\cdot{\bf b}\,-\,(\,-\,I\,)\cdot({\bf a}\times{\bf b}), \label{id-2}
\end{equation}
respectively. The geometrical meaning of the two identities is now transparent if we recall that the bivector
${(\,+\,I\cdot{\bf a})}$ represents a counterclockwise rotation about the ${\bf a}$-axis, whereas the bivector
${(\,-\,I\cdot{\bf a})}$ represents a clockwise rotation about the ${\bf a}$-axis. Accordingly, both identities interrelate
the points of a unit parallelized 3-sphere, but the identity (\ref{id-1}) interrelates points of a positively oriented
3-sphere whereas the identity (\ref{id-2}) interrelates points of a negatively oriented 3-sphere. In other words, the 3-sphere
represented by the identity (\ref{id-1}) is oriented in the counterclockwise sense, whereas the 3-sphere represented by the
identity (\ref{id-2}) is oriented in the clockwise sense. These two alternative orientations of the 3-sphere is then
the random hidden variable ${\lambda=\pm\,1}$ (or the initial state ${\lambda=\pm\,1}$) within my model. 

Given this geometrical picture, it is now easy to appreciate that identity (\ref{id-1}) corresponds to the physical space
characterized by the trivector ${\,+\,I\,}$, whereas identity (\ref{id-2}) corresponds to the physical space characterized
by the trivector ${\,-\,I\,}$ \ocite{Eberlein}. This is further supported by the evident fact that, apart from the choice of
a trivector, the identities (\ref{id-1}) and (\ref{id-2}) represent one and the same subalgebra. Moreover, there is clearly no
{\it a priori} reason for Nature to choose ${\,+\,I\,}$ as a fundamental trivector over ${\,-\,I\,}$. Either choice provides a
perfectly legitimate representation of the physical space, and neither is favored by Nature. Consequently, instead of
characterizing the physical space by fixed basis (\ref{cl30}), we can start out with two alternatively possible
characterizations of the physical space by the {\it hidden} basis
\begin{equation}
\left\{1,\,\;{\bf e}_x,\,{\bf e}_y,\,{\bf e}_z,\,\;{\bf e}_x\wedge{\bf e}_y,\,
{\bf e}_y\wedge{\bf e}_z,\,{\bf e}_z\wedge{\bf e}_x,\,\;\lambda\,
(\,{\bf e}_x\wedge{\bf e}_y\wedge{\bf e}_z\,)\right\}\!, \label{hid-cl30}
\end{equation}
where ${\lambda=\pm\,1}$. Although these considerations and the physical motivations behind them have been
the starting point of my program (see, for example, discussions in Refs.${\,}$\ocite{Christian-666},
\ocite{photon-666}, and \ocite{experiment-666}), Moldoveanu has overlooked them completely. In fact, as we
shall see in the next section, his arguments entirely depend on misidentifying the objective hidden variable\break
${\lambda}$ with the subjective choice of reference handedness to be made by the experimenter for computational purposes. 

Exploiting the natural freedom of choice in characterizing ${S^3}$ by either ${\,+\,I\,}$ or ${\,-\,I\,}$,
we can now combine the identities (\ref{id-1}) and (\ref{id-2}) into
a single hidden variable equation (at least for the computational purposes):
\begin{equation}
(\,\lambda\,I\cdot{\bf a})(\,\lambda\,I\cdot{\bf b})\,
=\,-\,{\bf a}\cdot{\bf b}\,-\,(\,\lambda\,I\,)\cdot({\bf a}\times{\bf b}), \label{combi-0}
\end{equation}
where ${\lambda=\pm\,1}$ now specifies the orientation of the 3-sphere. It is important to recognize that the difference
between the trivectors ${\,+\,I\,}$ and ${\,-\,I\,}$ in this equation
primarily reflects the difference in the handedness of the bivector
basis ${\{{\boldsymbol\beta}_x,\,{\boldsymbol\beta}_y,\,{\boldsymbol\beta}_z\}}$, and not in the handedness of the vector basis
${\{\,{\bf e}_x,\,{\bf e}_y,\,{\bf e}_z\}}$. This should be evident from the foregone arguments, but let us bring this point
home by considering the following change in the handedness of the vector basis:
\begin{equation}
+\,I\,=\,{{\bf e}_x}{{\bf e}_y}{{\bf e}_z}\;\;\longrightarrow\;\;(+\,{{\bf e}_x})(-\,{{\bf e}_y})(+\,{{\bf e}_z})
\,=\,-\,(\,{{\bf e}_x}{{\bf e}_y}{{\bf e}_z}\,)\,=\,-\,I\,.
\end{equation}
Such a change does not induce a change in the handedness of the bivector basis, since it leaves the product
${{\boldsymbol\beta}_x\,{\boldsymbol\beta}_y\,{\boldsymbol\beta}_z}$ unchanged. This can be easily
verified by recalling that ${{\boldsymbol\beta}_x\equiv I\cdot{\bf e}_x\,}$,
${\;{\boldsymbol\beta}_y\equiv I\cdot{\bf e}_y\,,\,}$ and
${\;{\boldsymbol\beta}_z\equiv I\cdot{\bf e}_z\,,\,}$ and consequently
\begin{equation}
+\,1\,=\,{\boldsymbol\beta}_x\,{\boldsymbol\beta}_y\,{\boldsymbol\beta}_z\;\;\longrightarrow\;\;
(\,-\,I\,)\cdot(+\,{{\bf e}_x})\,(\,-\,I\,)\cdot(-\,{{\bf e}_y})\,(\,-\,I\,)\cdot(+\,{{\bf e}_z})\,=\,
{\boldsymbol\beta}_x\,{\boldsymbol\beta}_y\,{\boldsymbol\beta}_z\,=\,+1\,.
\end{equation}
Conversely, a change in the handedness of bivector basis does not necessarily affect a change in the handedness of vector basis,
but leads instead to
\begin{equation}
+\,1\,=\,{\boldsymbol\beta}_x\,{\boldsymbol\beta}_y\,{\boldsymbol\beta}_z\;\;\longrightarrow\;\;
(\,-\,{\boldsymbol\beta}_x)\,(\,-\,{\boldsymbol\beta}_y)\,(\,-\,{\boldsymbol\beta}_z)\,=\,-\,
(\,{\boldsymbol\beta}_x\,{\boldsymbol\beta}_y\,{\boldsymbol\beta}_z\,)\,=\,-1\,,
\end{equation}
which in turn leads us back to equation (\ref{combi-0}) via equations (\ref{one-ident}) and (\ref{two-ident}). Thus the
sign difference between the trivectors ${\,+\,I\,}$ and ${\,-\,I\,}$ captured in equation (\ref{combi-0}) arises from the
sign difference in the product ${{\boldsymbol\beta}_x\,{\boldsymbol\beta}_y\,{\boldsymbol\beta}_z}$ and not from that in the
product ${{{\bf e}_x}{{\bf e}_y}{{\bf e}_z}}$. It is therefore of very different geometrical significance \ocite{Restoring}.
It corresponds to the difference between two possible orientations of the 3-sphere mentioned above. It is also important
to keep in mind that the combined equation (\ref{combi-0}) is simply a convenient shortcut for representing two completely
independent initial states of the system, one corresponding to the counterclockwise orientation of the 3-sphere and the
other corresponding to the clockwise orientation of the 3-sphere. Moreover, at no time these two alternative
possibilities are mixed during the course of an experiment. They represent two independent physical scenarios,
corresponding to two independent runs\break of the experiment. If we now use the notation ${{\boldsymbol\mu}=\lambda\,I}$,
then the combined identity (\ref{combi-0}) takes the convenient form
\begin{equation}
(\,{\boldsymbol\mu}\cdot{\bf a})(\,{\boldsymbol\mu}\cdot{\bf b})\,
=\,-\,{\bf a}\cdot{\bf b}\,-\,{\boldsymbol\mu}\cdot({\bf a}\times{\bf b}).\label{bi-identitzzzzzzzzzzzz}
\end{equation}

It should now be abundantly clear where the indefinite duality relation (\ref{1-on-1}) originates from. It is simply a convenient
shortcut describing the two alternate hidden variable possibilities encapsulated in this identity. Every
mathematician knows that what Bourbaki calls ``abuse of notation'' can, when handled with care, greatly illuminate what would
otherwise be a confusing situation. I do not claim that the indefinite duality relation (\ref{1-on-1}) is illuminating the
situation all that
much, but it does facilitate considerable ease in intricate computations \ocite{What-666}\ocite{illusion-666}. Thus Moldoveanu's
alarm about this relation is much ado about nothing. In any case, if one remains uncomfortable about using the identity
(\ref{bi-identitzzzzzzzzzzzz}), then there is always the option of working directly with the bivector basis themselves, as I
have done in my one-page paper \ocite{disproof}. Starting with equations (\ref{lisr}) and (\ref{lisr-left}) we first write the
basic hidden variable equation of the model as
\begin{equation}
{\boldsymbol\beta}_j\,{\boldsymbol\beta}_k\,=\,-\,\delta_{jk}\,-\,\lambda\;\epsilon_{jkl}\,
{\boldsymbol\beta}_l\,,
\end{equation}
with ${\lambda=\pm\,1}$ as a fair coin representing the two possible orientations of the 3-sphere. To see the equivalence
of this equation with the hidden variable identity (\ref{bi-identitzzzzzzzzzzzz}), let ${{\bf a}=a_{\! j}\,{\bf e}_j}$ and
${{\bf b}=b_{\! k}\,{\bf e}_k}$ be two unit vectors in ${{\rm I\!R}^3}$. We then have
\begin{equation}
\{\,\lambda\;a_j\;{\boldsymbol\beta}_j\,\}\,\{\,\lambda\;b_k\;{\boldsymbol\beta}_k\,\}\,=\,-\,a_j\,b_k\,\delta_{jk}\,
-\,\lambda\;\epsilon_{jkl}\;a_j\,b_k\;{\boldsymbol\beta}_l\,,
\end{equation}
which is equivalent to the identity (\ref{bi-identitzzzzzzzzzzzz}) with 
${(\,{\boldsymbol\mu}\cdot{\bf a})\equiv\{\,\lambda\;a_j\;{\boldsymbol\beta}_j\,\}}$,
${\;{\boldsymbol\mu}\cdot({\bf a}\times{\bf b})\equiv\{\,\lambda\;\epsilon_{jkl}\;a_j\,b_k\;{\boldsymbol\beta}_l\,\}}$, etc.
As a result, the standardized variables ${(\,{\boldsymbol\mu}\cdot{\bf a})\equiv\{\,\lambda\;a_j\;{\boldsymbol\beta}_j\,\}}$
and ${(\,{\boldsymbol\mu}\cdot{\bf b})\equiv\{\,\lambda\;b_k\;{\boldsymbol\beta}_k\,\}}$
immediately give rise to the EPR correlation: 
\begin{equation}
\lim_{n\,\gg\,1}\left[\frac{1}{n}\sum_{i\,=\,1}^{n}\,                                                                     
\left\{\,\lambda^i\;a_j\;{\boldsymbol\beta}_j\,\right\}\,\left\{\,\lambda^i\;b_k\;{\boldsymbol\beta}_k\,\right\}\right]
=\,-\,a_j\,b_j\,-\lim_{n\,\gg\,1}\left[\frac{1}{n}\sum_{i\,=\,1}^{n}\,                                                   
\left\{\,\lambda^i\,\epsilon_{jkl}\;a_j\,b_k\;{\boldsymbol\beta}_l\,\right\}\right]
=\,-\,a_j\,b_j\,+\,0\,=\,-\,{\bf a}\cdot{\bf b}\,, \label{corre-one}
\end{equation}
\begin{equation}
\text{and}\;\;\;\;\;\;\;\;\;\;\;\;\;\;\;\lim_{n\,\gg\,1}\left[\frac{1}{n}\sum_{i\,=\,1}^{n}\,
\left\{\,\lambda^i\,a_j\;{\boldsymbol\beta}_j\,\right\}\right]\,=\,0\,=\,
\lim_{n\,\gg\,1}\left[\frac{1}{n}\sum_{i\,=\,1}^{n}\,\left\{\,\lambda^i\,b_k\;{\boldsymbol\beta}_k\,\right\}\right].
\;\;\;\;\;\;\;\;\;\;\;\;\;\;\;\;\;\;\text{} \label{corre-two}
\end{equation}
It is important to remember that what is being summed over here are points of a parallelized 3-sphere representing
the outcomes of completely independent experimental runs in an EPR-Bohm experiment.
In statistical terms what these results are then showing is that correlation between the raw numbers
${{\mathscr A}({\bf a},\,{\boldsymbol\mu})\,=\,(-\,I\cdot{{\bf a}}\,)
\,(\,+\,{\boldsymbol\mu}\cdot{{\bf a}}\,)\,=\,\pm\,1\in S^3}$ and
${{\mathscr B}({\bf b},\,{\boldsymbol\mu})\,=\,(+\,I\cdot{{\bf b}}\,)
\,(\,+\,{\boldsymbol\mu}\cdot{{\bf b}}\,)\,=\,\pm\,1\in S^3}$
is ${-\,{\bf a}\cdot{\bf b}}$. According to Bell's theorem this is mathematically impossible. 
Further physical, mathematical, and statistical details of this ``impossible'' result can be found in
Refs.${\,}$\ocite{disproof} and \ocite{Restoring}.

\section{Handedness of the Bivectors is Not a Convention But an Initial EPR State}

In his preprint \ocite{M-2} Moldoveanu suggests that the above results are incorrect because I have used the
duality relation (\ref{1-on-1}) to derive them. That this claim is false should be clear from the above discussion, but
let us try to get to the heart of his misconception. This is revealed form the following statement he
makes, starting on the first page of \ocite{M-2}:
\begin{quote}
Even without spelling in detail the error, it is easy to see that the exterior product term should not vanish
on any handedness average because handedness is just a paper convention on how to consistently make computations.
For example one can apply the same incorrect argument to complex numbers because there is the same freedom to
choose the sign of ${\sqrt{-1}}$ based on the two dimensional coordinate handedness in this case. Then one can
compute the average of let's say ${z=3+2i}$ for a fair coin random distribution of handedness and arrive at the
incorrect answer: ${<\,z\,>\,=3}$ instead of ${<\,z\,>\,=z}$. 
\end{quote}
Now, to begin with, complex numbers do not have handedness. Thus his toy example badly misses the mark from the start.
In fact the problems with it go far deeper, but let us play along. If one treats
handedness as merely a ``paper convention''---say a choice between ${z=3+2i}$ and ${z=3-2i}$ for performing
computations---then of course one must maintain the same convention for all experimental runs to obtain the correct
answer, and that answer can only be ${<\,z\,>\,=z}$. However, in my model handedness of the bivectors basis, or more precisely
the orientation of the physical space ${S^3}$, is not a convention but a hidden variable ${\lambda}$. It is not a choice
that is made by an experimenter but by Nature herself, as an initial or complete state of a given pair of particles.
In other words, in my model the alternating handedness, ``${z=3+2i\,}$'' or ``${z=3-2i}$'', determine which
two of the four event detectors are triggered for a given pair of particles. Consequently the correct answer for the average
within my model cannot possibly be ${<\,z\,>\,=z}$ but ${<\,z\,>\,=3}$. Thus, indeed, even without spelling
out his error in detail we can see where Moldoveanu has gone wrong.

One might think that it is the unconventional nature of my framework that might have misled Moldoveanu
into making such an elementary mistake. Reading his preprints suggests otherwise, however.
For instance, his argument seems to overlook the fact that, since each specific instance of the hidden
variable ${\lambda}$---i.e., each specific initial orientation of the 3-sphere---specifies an initial state of the
EPR pair in my model, it corresponds to a physical scenario quite independent of the previous or subsequent
initial state of the pair. Thus, as in Bell's own local model \ocite{Bell-Reply},
Alice makes a series of measurements of {\it different} particles, not
repeated measurements of the same particle.
Each particle thus begins with a different initial state (or a different
common cause) ${\lambda}$, which interacts with Alice's analyzer through the bivector
${\{\,a_j\;{\boldsymbol\beta}_j(\lambda)\,\}}$ to produce her measurement outcome ${{\mathscr A}({\bf a},\,\lambda)}$.
No mixing of algebra of any kind occurs during this process since what is averaged over are real numbers observed
independently by Alice and Bob.

\section{Failure of Bell's Theorem Implies Vindication of the EPR Argument}

One of several misreadings of my work by Moldoveanu is reflected in his claim that my counterexample to Bell's theorem
is based on assumptions different from those of Bell. To see that this claim is false, let us recall the
bare essentials of Bell's assumptions. These appear no more starkly than in Bell's own replies to his
critics \ocite{Bell-Reply}. Bell insists that it is {\it not} possible to find local functions of the form
\begin{align}
{\mathscr A}({\bf a},\,{\lambda})\,&=\,+\,1\;\,\;\text{or}\;\,-1 \label{1-new} \\
\text{and}\;\;{\mathscr B}({\bf b},\,{\lambda})\,&=\,+\,1\;\,\;\text{or}\;\,-1\, \label{2-new}
\end{align}
which can give the correlation of the form
\begin{equation}
\langle\,{\mathscr A}{\mathscr B}\;\rangle\,=\,-\,{\bf a}\cdot{\bf b}\,, \label{3-new}
\end{equation}
where the measurement setting ${\bf b}$ of a particular polarizer has no effect on what happens, ${\mathscr A}$, in a remote region,
and likewise that the measurement setting ${\bf a}$ has no effect on ${\mathscr B}$. ``{\it This is the theorem}'',
he insists (my emphasis).

Now the first thing to recall here is that if this theorem is false, then there is nothing to impede the conclusion by
EPR: {\it the description of physical reality provided by quantum mechanics is incomplete}. Moreover, demonstration of
incompleteness of any theory for a single physical scenario is sufficient to demonstrate the incompleteness
of that theory for {\it all} physical scenarios. Thus failure of Bell's theorem as stated above necessitates incompleteness
of quantum mechanics as a whole \ocite{illusion-666}. It seems, however, that Moldoveanu has not appreciated the force of
this logic. All other ``impossibility proofs'', no matter how elaborate, are powerless against the logic of the EPR argument.
Consequently, irrespective of the arguments of the previous sections, my one-page paper \ocite{disproof} by itself is more
than sufficient to vindicate EPR, repudiate Bell, and show that quantum mechanics is necessarily an incomplete theory of nature.
This is because the variables ${{\mathscr A}({\bf a},\,{\lambda})=\pm\,1}$ and ${{\mathscr B}({\bf b},\,{\lambda})=\pm\,1}$
defined in equations (1) and (2) of that paper, together with the correlation between them obtained in equation (7),
decisively and unambiguously contradict the assertion made by Bell in equations (\ref{1-new}) to (\ref{3-new}) above. What
is more, they do so {\it irrespective of the interpretation attached to the mathematical terms involved in the calculation
of the correlation in} \ocite{disproof}. It is therefore quite puzzling how anyone who has studied my paper and understood its
logic (as well as that of EPR) can continue to believe in Bell's theorem.

\section{The Local-Realistic Framework in Question is Strictly Non-Contextual}

Now the topological underpinning of my model is quite different from that of usually considered
local hidden variable theories, which are implicitly
expected to be contextual in general. Usually one expects the numbers ${\mathscr A}$ and ${\mathscr B}$
to change when the directions of measurements ${\bf a}$ and ${\bf b}$ are changed. This informal expectation, however, is profoundly
misguided. No such local contextual change is ever observed in the actual experiments, or even predicted by quantum mechanics.
Locally one only finds purely random measurement outcomes with no hidden order within them. The variations that are predicted and
observed are only in the correlation between the numbers ${\mathscr A}$ and ${\mathscr B}$, and not in the randomness
of the numbers ${\mathscr A}$ and ${\mathscr B}$ themselves \ocite{Nature-666}.
And that is precisely what is predicted by my model.   

Nevertheless, despite the fact that my framework is manifestly non-contextual, Moldoveanu has claimed that it is contextual.
To appreciate the evident falsity of this claim, let us have a closer look at the relationship between the statistically
pertinent standard scores ${{\boldsymbol\mu}\cdot{\bf a}}$ and ${{\boldsymbol\mu}\cdot{\bf b}}$ and the actually observed raw
scores ${{\mathscr A}({\bf a},\,{\boldsymbol\mu})}$ and ${{\mathscr B}({\bf b},\,{\boldsymbol\mu})}$:
\begin{equation}
S^3\ni {\mathscr A}({\bf a},\,{\boldsymbol\mu})\,=\,(-\,I\cdot{{\bf a}}\,)
\,(\,+\,{\boldsymbol\mu}\cdot{{\bf a}}\,)\,=\,
\begin{cases}
+\,1\;\;\;\;\;{\rm if} &{\boldsymbol\mu}\,=\,+\,I \\
-\,1\;\;\;\;\;{\rm if} &{\boldsymbol\mu}\,=\,-\,I
\end{cases} \label{17-Joy}
\end{equation}
and
\begin{equation}
S^3\ni {\mathscr B}({\bf b},\,{\boldsymbol\mu})\,=\,(+\,I\cdot{{\bf b}}\,)
\,(\,+\,{\boldsymbol\mu}\cdot{{\bf b}}\,)\,=\,
\begin{cases}
-\,1\;\;\;\;\;{\rm if} &{\boldsymbol\mu}\,=\,+\,I \\
+\,1\;\;\;\;\;{\rm if} &{\boldsymbol\mu}\,=\,-\,I\,,
\end{cases} \label{18-Joy}
\end{equation}
with equal probabilities for ${\boldsymbol\mu}$ being either ${+\,I}$ or ${-\,I}$.
Note that ${{\mathscr A}({\bf a},\,{\boldsymbol\mu})}$ and ${{\mathscr B}({\bf b},\,{\boldsymbol\mu})}$,
in addition to being manifestly realistic, are strictly {\it local} variables. Moreover, it is not difficult
to see that they are manifestly non-contextual \ocite{Contextual}. 
Alice's measurement result, although refers to her freely chosen context ${\bf a}$, depends only on the
initial state ${{\boldsymbol\mu}}$; and likewise, Bob's measurement result, although refers to his freely chosen context
${\bf b}$, depends only on the initial state ${{\boldsymbol\mu}\,}$. In other words, all possible measurement results at all
possible angles are determined entirely by the initial orientation of the 3-sphere specified by ${{\boldsymbol\mu}}$,
and do not change when the local contexts are changed. This fact is so manifestly obvious from the above definitions of
${{\mathscr A}({\bf a},\,{\boldsymbol\mu})}$ and ${{\mathscr B}({\bf b},\,{\boldsymbol\mu})}$ that it makes one wonder
why anyone would think my framework is contextual.
Could it be because the standard scores ${{\boldsymbol\mu}\cdot{\bf a}}$ and ${{\boldsymbol\mu}\cdot{\bf b}}$
somehow appear to be contextual? It is easy to check, however, that they are not:
\begin{equation}
{\boldsymbol\mu}\cdot{\bf n}\,=
\begin{cases}
+\,1\;\,{\rm about}\,\,{\bf n} & \text{${\;\;\;}$if ${\,{\boldsymbol\mu}\,=\,+\,I}$,} \\
-\,1\;\,{\rm about}\,\,{\bf n} & \text{${\;\;\;}$if ${\,{\boldsymbol\mu}\,=\,-\,I}$.} \label{haha666}
\end{cases}
\end{equation}
Evidently, the values of the standard scores  ${{\boldsymbol\mu}\cdot{\bf n}}$ also do not depend upon the experimental
context ${\bf n}$. For instance, if the context is changed from ${\bf n}$ to ${\bf n'}$, we obtain
\begin{equation}
{\boldsymbol\mu}\cdot{\bf n'}\,=
\begin{cases}
+\,1\;\,{\rm about}\,\,{\bf n'} & \text{${\;\;\;}$if ${\,{\boldsymbol\mu}\,=\,+\,I}$,} \\
-\,1\;\,{\rm about}\,\,{\bf n'} & \text{${\;\;\;}$if ${\,{\boldsymbol\mu}\,=\,-\,I}$,} \label{nhanha666}
\end{cases}
\end{equation}
and the corresponding raw scores remain exactly the same. This is not surprising, because, as stressed above, according to
my model all measurement results are simply detecting the orientation of the 3-sphere specified by the initial state
${\,{\boldsymbol\mu}}$
(recall also that an answer to any quantum mechanical question can always be reduced to a set of Yes/No answers).

So, clearly, as far as the EPR-Bohm correlations are concerned, my model of the physical reality is not contextual. 
But what about more general quantum correlations? Could my local-realistic framework be contextual for general quantum
correlations? Perhaps that is what Moldoveanu is hoping for \ocite{M-1}. It is not difficult to see however that my framework is
manifestly non-contextual even for the most general case. This fact can be stated as a theorem:

\vspace{0.35cm}
{\parindent 0pt
\underbar{\bf Theorema Egregium:}
\vspace{0.2cm}
\begin{itemize}
\item[${}$]Every quantum mechanical correlation among a set of measurement results, such as ${{\mathscr A}=\pm\,1}$,
${{\mathscr B}=\pm\,1}$, ${{\mathscr C}=\pm\,1}$, etc.,${\;}$can be understood as a
classical, local-realistic correlation among a set of points of a parallelized 7-sphere.
\end{itemize}
The proof of this theorem (or at least a sketch of it) can be found in section IVA of Ref.${\,}$\ocite{What-666}
and section VI of Ref.${\,}$\ocite{illusion-666}.}

Now 7-sphere has a very rich topological structure. It happens to be homeomorphic to the space of unit octonions, which are
well known to form the most general possible division algebra. In the language of fiber bundles one can view 7-sphere as a
4-sphere worth of 3-spheres. Each fiber of the 7-sphere is then a 3-sphere, and each one of these 3-spheres is a 2-sphere
worth of circles. Thus the four parallelizable spheres---${S^0}$, ${S^1}$, ${S^3}$, and ${S^7}$---can all be viewed as nested
within a 7-sphere. The EPR-Bohm correlations can then be understood as correlations among the equatorial points of one of the
fibers of this 7-sphere, as we saw in section II. Alternatively, the 7-sphere can be thought of as a 6-sphere worth of circles.
Thus the above theorem can be framed entirely in terms of circles, each one of which described by a classical, octonionic spinor
with a well-defined sense of rotation (i.e., whether it describes a clockwise rotation about a point within the 7-sphere or a
counterclockwise rotation). This sense of rotation in turn defines a definite handedness (or orientation) about every point
of the 7-sphere. If we designate this handedness by a random number ${\lambda}$, then local measurement results for any
physical scenario can be represented by the raw scores of the form
\begin{equation}
S^7\ni {\mathscr A}({\bf a},\,{\boldsymbol\mu})\,=\,(-\,J\cdot{{\bf N(a)}}\,)\,(\,+\,{\boldsymbol\mu}\cdot{{\bf N(a)}}\,)\,=\,
\begin{cases}
+\,1\;\;\;\;\;{\rm if} &{\boldsymbol\mu}\,=\,+\,J \\
-\,1\;\;\;\;\;{\rm if} &{\boldsymbol\mu}\,=\,-\,J\,,
\end{cases} \label{17876-Joy}
\end{equation}
where ${{\bf a}\in{\rm I\!R}^3}$ and ${{\bf N(a)}\in{\rm I\!R}^7}$ are unit vectors and ${{\boldsymbol\mu}=\lambda\,J}$ is the
hidden variable analogous to ${{\boldsymbol\mu}=\lambda\,I}$ with ${I={\bf e}_x}{{\bf e}_y}{{\bf e}_z}$ replaced by
\begin{equation}
J\,=\,{{\bf e}_1}{{\bf e}_2}{{\bf e}_4}\,+\,{{\bf e}_2}{{\bf e}_3}{{\bf e}_5}\,+\,{{\bf e}_3}{{\bf e}_4}{{\bf e}_6}\,+
\,{{\bf e}_4}{{\bf e}_5}{{\bf e}_7}\,+\,{{\bf e}_5}{{\bf e}_6}{{\bf e}_1}\,+\,{{\bf e}_6}{{\bf e}_7}{{\bf e}_2}\,+
\,{{\bf e}_7}{{\bf e}_1}{{\bf e}_3}\,.\label{chosen-try}
\end{equation}
The standard scores corresponding to the raw scores (\ref{17876-Joy}) are then given by ${{\boldsymbol\mu}\cdot{\bf N(a)}}$, which
geometrically represent the equatorial points of a parallelized 7-sphere, just as 
${{\boldsymbol\mu}\cdot{\bf a}}$ represented the equatorial points of a parallelized 3-sphere.

It is now clear that even in the most general case the variables representing measurement results in my framework
are manifestly non-contextual.
Just as before, Alice's measurement result, although refers to her freely chosen context
${\bf a}$, depends only on the initial state ${{\boldsymbol\mu}=\lambda\,J}$; and likewise, Bob's measurement result
may refer to his freely chosen context ${\bf b}$, but would depend only on the initial state ${{\boldsymbol\mu}=\lambda\,J}$.
In other words, all possible
measurement results at all possible angles are determined entirely by the initial orientation of the 7-sphere specified by
${{\boldsymbol\mu}=\lambda\,J}$,
and do not change when the local contexts are changed${\,}$\footnote{Often
in physics formal expositions end up obscuring the simplest of truths.
Let me therefore try to redraw the above picture in homely terms of the science fiction novel Flatland \ocite{Flatland}.
Suppose we have proof that we are living in a parallelized 3-sphere rather than ${{\rm I\!R}^3}$. Then it would not
surprise us that correlations between certain random binary events turn out to be sinusoidal rather than linear,
because that is what the topology of the 3-sphere dictates \ocite{Restoring}\ocite{experiment-666}.
More generally, if we had proof that we
were living in a parallelized 7-sphere rather than ${{\rm I\!R}^3}$, then it would not surprise us that correlations
between certain random binary events turn out to be stronger than linear, because that is what the topology of the 7-sphere
dictates \ocite{What-666}\ocite{illusion-666}. We would then not worry about contextuality or non-locality, but only
about the topology of the 7-sphere. But that is precisely what my framework is suggesting. It is suggesting that quantum
correlations are the evidence, not of non-locality or non-reality of any kind, but of the fact that we are
living in a parallelized 7-sphere.}. Moreover, the values of even the standard scores
${{\boldsymbol\mu}\cdot{\bf N(a)}}$ do not change when the local contexts are changed. To be sure, when the 3D context is
changed, say from ${\bf a}$ to ${\bf a'}$, the corresponding 7D direction changes from ${\bf N}$ to ${\bf N'}$, but that does
not at all affect the values of either the raw scores or the standard scores, because they are determined entirely by the
initial orientation of the 7-sphere specified by ${{\boldsymbol\mu}=\lambda\,J}$.

\section{Non-Commuting Standard Scores are Merely Calculational Tools}

It is clear from the above discussion that my entire local-realistic framework is strictly non-contextual. Nevertheless,
in his preprint \ocite{M-1} Moldoveanu has argued that it somehow ``must be'' contextual. His argument relies on some
well known (but irrelevant) theorems (see below) and the non-commutativity of the bivectors, which, as we saw, plays the
role of standard scores within my framework \ocite{Restoring}. What is amiss in his argument, however, is the evident fact
that non-commutativity enters in my framework only at the level of standard scores, not raw scores. In fact in either of
his preprints there is no appreciation of the vital conceptual difference between the raw scores and standard scores, let
alone the significance of this difference within my framework. Had he appreciated this conceptual difference (as explained,
for example, in Ref.${\,}$\ocite{Restoring}), he would have recognized that non-commutativity of the standard scores
within my framework---which he claims makes contextuality inevitable in general---is only an intermediate calculational tool.
The actual eventualities, ${\mathscr A}$, ${\mathscr B}$, etc., ({\it i.e.}, the actual measurement results) always commute
with each other,
\begin{equation}
[\,{\mathscr A},\,{\mathscr B}\,]\,=\,0\,,\;\;\;\forall\;\,{\bf a},\,\;{\bf b},\,\;\text{and}\,\;\lambda\,,
\end{equation}
because they are simply scalar numbers ({\it cf.} their definition (\ref{17876-Joy}) above). In statistical terms,
these measurement results are raw scores, and the corresponding non-commuting variables---i.e., the bivectors
${{\boldsymbol\mu}\cdot{\bf N(a)}}$---are standard scores. The standard scores---or the standardized variables---indeed
do not commute in general, but they are simply intermediate calculational tools, not something that is actually observed
in the experiments.
Therefore Moldoveanu is quite mistaken in building a case around the non-commutativity of such mathematical
constructs. Moreover, the mystique of classical non-commutativity within my model completely evaporates when one
notes that it can always be understood as a vector addition in a higher-dimensional space (see, for example, discussion
below Eq.${\,}$(38) in Ref.${\,}$\ocite{illusion-666}).\break
Thus non-commutativity within my model does not have the ontological significance Moldoveanu thinks it has.

Leaving aside this statistical misconception, let us see whether Moldoveanu's
argument for the contextuality within my framework itself holds water. In fact
his argument turns out to be both logically and conceptually incongruent even if we accept its
premises (which are based on multiple misconceptions of my framework in any case).
To be more specific, his argument relies on theorems against non-contextual hidden variable theories such as that
by Kochen and Specker and some of its lesser known extensions, but without spelling out how exactly
such theorems---based on discrete spaces of measurement results
as they are---are applicable to my framework based on topology and continuity.
In fact they are not at all applicable, because none of them even remotely address the topological concerns I have raised
within the context of Bell's theorem and its variants \ocite{What-666}\ocite{illusion-666}.
Thus his attempt of trying to fit my topological framework within a preconceived conceptual box of contextuality is like
trying to fit a square peg in a round hole. 

\section{A Simply-connected Model Cannot be Simulated by its Discretized Imitation}

As we saw above, within my framework quantum correlations are explained as topological effects, not contextual
effects, and that Moldoveanu has overlooked this obvious fact. This is reflected, for example, in his failed
attempts to simulate my model on a computer by expecting some sort of contextual variation in the measurement
results. The usual idea behind a computer simulation of EPR correlations is to program how a measurement function,
say ${{\mathscr A}({\bf a},\,\lambda)}$, changes when its context is changed, say from ${\bf a}$ to ${\bf a'}$,
and likewise for the function ${{\mathscr B}({\bf b},\,\lambda)}$. But in my model these
functions do not change with their contexts at all. It is the topology of the physical space that brings about
the sinusoidal correlation between ${{\mathscr A}({\bf a},\,\lambda)}$ and  ${{\mathscr B}({\bf b},\,\lambda)}$,
not the contextual variations within ${{\mathscr A}({\bf a},\,\lambda)}$ and  ${{\mathscr B}({\bf b},\,\lambda)}$
themselves. As counterintuitive as this may seem, that is what the mathematics of my model implies, and it matches\break
exactly with the experimental evidence. Consequently, most unsophisticated simulation attempts are bound to fail. 
 
In fact, quite independently of Moldoveanu's failed attempts, in my view the whole fashion of simulating EPR correlations
on a computer is completely wrong headed. It is based on serious misconceptions about the true physical and mathematical
reasons for the existence of EPR correlations in Nature \ocite{What-666}. In all real-life demonstrations of the correlations,
Alice and Bob are known to always observe truly random outcomes of their measurements: ${{\mathscr A}=\pm\,1}$ and
${{\mathscr B}=\pm\,1}$ \ocite{Nature-666}. Therefore, as correctly recognized by Bell, no local functions of the form
${{\mathscr A}({\bf a},\,\lambda)}$ and ${{\mathscr B}({\bf b},\,\lambda)}$ can reproduce the observed correlation,
{\it unless the topological properties of the physical space itself are taken into account}.
In the language of my model this means that one must first model the physical space, not as
${{\rm I\!R}^3}$, but as ${S^3}$, which differs from ${{\rm I\!R}^3}$ only by a single point at infinity \ocite{Restoring}.
By contrast, what is usually tried in attempts to simulate my model is a completely wrong headed approach,
based on an implicit assumption that the numbers ${{\mathscr A}}$ and ${{\mathscr B}}$ are only 
apparently but not truly random, and if only one can somehow discover the correct functional dependence of these
numbers on the disposition of the apparatus and hidden variables then the correct correlation between them would emerge.
However, as Bell convincingly demonstrated long ago \ocite{Bell-1964-666},
one can never reproduce the sinusoidal correlation in this manner. For
the EPR correlation are what they are because of the topological properties of the physical space itself \ocite{experiment-666},
{\it not} because there exists some as-yet-uncovered hidden order in the randomness of ${{\mathscr A}}$ and ${{\mathscr B}}$.
Moldoveanu would have saved himself a lot of time and effort had he appreciated this basic message of my framework.

In any case, a simply-connected model such as mine cannot possibly be either proved or disproved by its numerical simulation.
A simulation of a model is an implementation of its analytical details, not an experiment that can either prove
or disprove its validity. If reality can be so simply simulated then there would be no need for the staggeringly
expensive actual experiments. Reality is mathematically far richer and profounder than what a computer can fathom.

\section{Counting the Number of Trivectors has no Bearing on the Integration}

In his preprint \ocite{M-2} Moldoveanu makes a claim that ``Isotropically weighted averages
of non-scalar part of correlations and measurement outcomes cannot be both zero.'' To justify
this claim he then enters into some strange counting of the ${{\boldsymbol\mu}}$'${\rm s}$
in the integrations (18) and (19) of my primary paper \ocite{Christian-666}. He concludes
from this counting that evaluation of both of these integrations cannot be right. However,
while counting the ${{\boldsymbol\mu}}$'${\rm s}$ he does not seem to realize that there is no
operational difference between the non-scalar part of correlations and measurement outcomes.
The non-scalar part of the correlation is just another possible measurement outcome, albeit
along the exclusive direction ${{\bf a}\times{\bf b}}$. He also fails to take into account
the obvious fact that the vector manifold in question over which
the integrations are performed is related to the 3-sphere, which has a highly non-trivial
topology reflected in the identity (17) of my paper. Moreover, he fails to recognize
that equations (18) and (19) involve integrations over a random variable distributed
over this non-trivial topology, and therefore a simple counting of
the ${{\boldsymbol\mu}}$'${\rm s}$ cannot possibly yield any insight into what the actual
evaluation of a given integration would yield. In any case, for his conclusions he
relies on one of his previous arguments which I have already refuted in section II above. 
Thus, in the light of the explicit and unambiguous results (\ref{corre-one}) and
(\ref{corre-two}) obtained above, his intended argument here has neither any relevance
nor any meaning for my model.

\section{Twist in the Clifford Parallels Cannot be Revealed by an Incorrect Rotor}

By now it is quite evident that much of Moldoveanu's discussion has nothing whatsoever to do with my model. A further
illustration of this fact is his false claim that in Ref.${\,}$\ocite{Restoring} I have used an incorrect rotor to parallel
transport a ``bivector'' (in fact I parallel transport a multivector, but never mind). What he fails to acknowledge, however,
is that the parallel transport in question is part of a heuristic demonstration of an already proved result. By this stage
in Ref.${\,}$\ocite{Restoring} I have rigorously derived
the EPR correlation (as in equation (\ref{corre-one}) above), and already demonstrated how the correct combination
of measurement outcomes, namely ${++}$, ${--}$, ${+-}$, and ${-+}$, arises due to a twist in the fibration of the
3-sphere, which is taken in my model as the physical space. The purpose of the heuristic demonstration is then to provide
an additional intuitive understanding of how this twist brings about the strong quantum correlation. 

The rotor I have used for this demonstration (with tildes on the vectors dropped to simplify notation) is of the form
\begin{equation}
{\bf a}\,{\bf b}\,=\,{\cal R}_{{{\bf a}}{{\bf b}}}\,=\,\exp\left\{(\,I\cdot{{\bf c}}\,)\,
\phi_{{{\bf a}}{{\bf b}}}\right\}\,,\;\;\;\;\text{with}\;\;
{{\bf c}}:=\frac{{{\bf a}}\times{{\bf b}}}{|{{\bf a}}\times{{\bf b}}|}
\,\equiv\,\frac{{{\bf a}}\times{{\bf a'}}}{|{{\bf a}}\times{{\bf a'}}|}
\,\equiv\,\frac{{{\bf b}}\times{{\bf b'}}}{|{{\bf b}}\times{{\bf b'}}|}\,, \label{rotrotrot}
\end{equation}
where the respective angles ${\psi_{{{\bf a}}{{{\bf a'}}}}}$ and ${\psi_{{{\bf b}}{{{\bf b'}}}}}$
between ${{\bf a}}$ and ${{{\bf a'}}}$ and ${{\bf b}}$ and ${{{\bf b'}}}$ are assumed to be
infinitesimally small. The reason for this particular choice of rotor has to do with the
fact that it happens to be the correct choice to illustrate\break the twist in the Hopf fibration of the 3-sphere \ocite{Eguchi}.
It is well known that this twist can be quantified by the relation
\begin{equation}
e^{i\psi_b}\,=\,e^{i\phi}\,e^{i\psi_a}\,, \label{72-3}
\end{equation}
where ${\psi_a}$ and ${\psi_b}$ are fiber coordinates above the two hemispheres of a 2-sphere (taken as the base manifold),
${\phi}$ is an angle parameterizing a thin strip around its equator, and ${e^{i\phi}}$ is the
transition function gluing the two sections into a full 3-sphere \ocite{Eguchi}. It is easy to see from this equation
that the fiber coordinates match perfectly at the angle ${\phi=0}$ (modulo
${2\pi}$), but differ from each other at all other intermediate angles. For instance,
${e^{i\psi_a}}$ and ${e^{i\psi_b}}$ differ by a minus sign at ${\phi=\pi}$.
Now in the coordinate-free language of geometric algebra the above relation can be expressed as
\begin{equation}
{\bf b}\,{\bf b'}\,=\,{\bf a}\,{\bf b}\;{\bf a}\,{\bf a'}
\end{equation}
for all ${\bf a}$'s and ${\bf b}$'s ${\in\,{\rm I\!R}^3}$,
provided we identify the infinitesimal angles ${\psi_{{{\bf a}}{{{\bf a'}}}}}$ and ${\psi_{{{\bf b}}{{{\bf b'}}}}}$ between
${{\bf a}}$ and ${{{\bf a'}}}$ and ${{\bf b}}$ and ${{{\bf b'}}}$ with the fiber coordinates ${\psi_a}$ and ${\psi_b\,}$, and
the finite angle between ${{\bf a}}$ and ${{{\bf b}}}$ with the generator of the transition function ${e^{i\phi}}$.
In the language of my model this generalized relation is then equivalent to equation (45) of Ref.${\,}$\ocite{Restoring}:
\begin{equation}
(+\,I\cdot{{\bf b}}\,)\,(\,+\,{\boldsymbol\mu}\cdot{{\bf b'}}\,)\,=\,
{\cal R}_{{{\bf a}}{{\bf b}}}\,\left\{(\,+\,I\cdot{{\bf a}}\,)\,(\,+\,{\boldsymbol\mu}\cdot{{\bf a'}}\,)\right\}.
\end{equation}

It is now clear that the rotor I have used in Ref.${\,}$\ocite{Restoring} is {\it the} correct rotor for the
purpose at hand. The rotor Moldoveanu has considered, on the other hand, has no relevance either for my model or for the
demonstration of the twist
under consideration. To justify his choice he goes on to produce arguments which further reveal
his lack of appreciation, not only of the topology of the 3-sphere, but also of some of the most basic facts
about geometric algebra. For instance he repeatedly identifies the quantity
${[(+\,I\cdot{{\bf a}}\,)\,(\,+\,{\boldsymbol\mu}\cdot{{\bf a'}}\,)]}$ as ``bivector'' when it is evidently a multivector
(in fact a quaternion or a rotor). Consequently, his subsequent discussion is marred by nonsensical statements like
``...the bivectors ${[(+\,I\cdot{{\bf a}}\,)\,(\,+\,{\boldsymbol\mu}\cdot{{\bf a'}}\,)]}$ and
${[(+\,I\cdot{{\bf b}}\,)\,(\,+\,{\boldsymbol\mu}\cdot{{\bf b'}}\,)]}$ are actually identical because they have the same
orientation, magnitude, and sense of rotation.'' But they are not \ocite{Restoring}\ocite{What-666}.
The two quaternions in question in fact represent two
entirely different points of the 3-sphere \ocite{Restoring}. These two points are anything but identical.
What is more, because he thinks that ${[(+\,I\cdot{{\bf a}}\,)\,(\,+\,{\boldsymbol\mu}\cdot{{\bf a'}}\,)]}$ is a ``bivector''
he suggests a strange direction about which I should have parallel \break
transported my ``bivector.'' But his suggested direction has no relevance either for my model or for the 3-sphere.

\section{Null Bivector is a Bivector that Subtends No Meaningful Area}

As we just noted, in Ref.${\,}$\ocite{Restoring}, in the course of the demonstration discussed above,
I consider a bivector of the form
\begin{equation}
I\cdot{{\bf c}}\,,\;\;\;\;\;\text{with}\;\;\;
{{\bf c}}:=\frac{{{\bf a}}\times{{\bf a}\,'}}{|{{\bf a}}\times{{\bf a}\,'}|}\,,
\end{equation}
and state that in the limit ${{{\bf a}\,'}\rightarrow{{\bf a}}}$ the bivector
${I\cdot{{\bf c}}}$ reduces to a null bivector.
Moldoveanu disputes this statement and produces a lengthy and convoluted argument to claim that the limit operation
involved here is mathematically illegal. My statement, however, is trivially correct, and there is no illegal limiting
operation of any kind involved in my reasoning. To see how trivial the issue is and how mistaken his argument is,
let us define the following two vectors:
\begin{equation}
{\widetilde{\bf a}}\,=\frac{{{\bf a}}}{\sqrt{|{{\bf a}}\times{{\bf a}\,'}|}\;}\;\;\;\;\;\text{and}
\;\;\;\;\;{\widetilde{\bf a}\,'}=\frac{{{\bf a}\,'}}{\sqrt{|{{\bf a}}\times{{\bf a}\,'}|}\;}\,.
\end{equation}
We can now rewrite the bivector ${I\cdot{{\bf c}}}$ as
\begin{equation}
I\cdot{{\bf c}}\,=\,I\cdot(\,{{\widetilde{\bf a}}\times{\widetilde{\bf a}\,'}})
\,=\,{{\widetilde{\bf a}}\wedge{\widetilde{\bf a}\,'}}.
\end{equation}
Now it is true that in the limit ${{{\bf a}\,'}\rightarrow{{\bf a}}}$ not only the directions of the vectors
${{\widetilde{\bf a}}}$ and ${{\widetilde{\bf a}\,'}}$ tend to coincide but also their lengths tend to infinity.
The question then is, whether or not the following statement is true:
\begin{equation}
\lim_{{{\bf a}\,'}\rightarrow\,{{\bf a}}}\;
{{\widetilde{\bf a}}\wedge{\widetilde{\bf a}\,'}}\,=\,0\,.
\end{equation}
The answer is that it is trivially true. Despite the fact that in the limit ${{{\bf a}\,'}\rightarrow{{\bf a}}}$ the lengths
of the vectors ${{\widetilde{\bf a}}}$ and ${{\widetilde{\bf a}\,'}}$ tend to infinity, there can be no meaningful area spanned
by the resulting vector ${{\widetilde{\bf a}}={\widetilde{\bf a}\,'}}$, and hence what emerges in the limit is a null bivector.
This is a standard understanding of null bivector found in any textbook \ocite{Clifford}. Just as a null vector is a vector that
has no meaningful length, a null bivector is a bivector that has no meaningful area.

But let us not rely on the authority of good books on geometric algebra. Let us instead see the absurdity
of Moldoveanu's argument for ourselves. His claim---supported by lengthy and
elaborate arguments---is that the bivector ${{{\widetilde{\bf a}}\wedge{\widetilde{\bf a}\,'}}}$ remains
non-null even in the limit ${{{\bf a}\,'}\rightarrow{{\bf a}}}$. In other words, his argument is that the bivector
${{{\widetilde{\bf a}}\wedge{\widetilde{\bf a}\,'}}}$ remains non-null even when the vectors ${{\widetilde{\bf a}}}$ and
${{\widetilde{\bf a}\,'}}$ are one and the same vector and the area spanned by it is zero. This is as
absurd as claiming that a vector remains non-null even when the distance between its end points is zero.

In sum, because of various misconceptions about the basic concepts in geometric algebra, many of the categorical
assertions made by Moldoveanu in his preprint \ocite{M-2} are simply wrong. In particular, there is nothing wrong
with my statement in Ref.${\,}$\ocite{Restoring} that ${I\cdot{{\bf c}}}$ reduces to a null bivector in the limit
${{{\bf a}\,'}\rightarrow{{\bf a}}}$. Consequently, my argument starting from Eq.${\,}$(42) and ending after
Eq.${\,}$(46) in that paper is an entirely cogent argument, illustrating how the illusion of quantum non-locality
arises in the EPR case due to a twist in the Hopf fibration of 3-sphere. Could Moldoveanu's failure to appreciate
this simple illustration be due to his failure to appreciate the topology of the 3-sphere itself?

\section{Correlation Between Raw Scores is given by Covariance between Standard Scores}

At several junctures in the preprint \ocite{M-2} Moldoveanu reasons as follows: Since the outcomes of Alice and Bob are
``always the same'' for all directions according the definitions (\ref{17-Joy}) and (\ref{18-Joy}) above, it follows
that the product of these outcomes, and hence the correlation between them, will always be equal to ${+\,1}$ regardless
of the measurement directions chosen by Alice and Bob, contradicting the prediction ${-\,{\bf a}\cdot{\bf b}}$
of quantum mechanics. This conclusion, which is of course false on several counts, stems from a failure to appreciate
the basic rules of statistical inference, not to mention the basic topology of the 3-sphere \ocite{Restoring}\ocite{What-666}.
More specifically, what Moldoveanu has failed to recognize is that
${{\mathscr A}({\bf a},\,{\boldsymbol\mu})}$ and ${{\mathscr B}({\bf b},\,{\boldsymbol\mu})}$
are generated with {\it different}
bivectorial scales of dispersion, and
hence the correct correlation between them can be inferred only by calculating the covariation of the
corresponding standard scores ${{\boldsymbol\mu}\cdot{\bf a}}$ and ${{\boldsymbol\mu}\cdot{\bf b}\,}$:
\begin{align}
{\cal E}({\bf a},\,{\bf b})\,=\lim_{\,n\,\gg\,1}\left[\frac{1}{n}\sum_{i\,=\,1}^{n}\,
{\mathscr A}({\bf a},\,{\boldsymbol\mu}^i)\;{\mathscr B}({\bf b},\,{\boldsymbol\mu}^i)\right]
\,=\lim_{\,n\,\gg\,1}\left[\frac{1}{n}\sum_{i\,=\,1}^{n}\,
(\,{\boldsymbol\mu}^i\cdot{\bf a})(\,{\boldsymbol\mu}^i\cdot{\bf b})\right]
\,=\,-\,{\bf a}\cdot{\bf b}\,.
\end{align}
I have explained the relationship between raw scores and standard scores in great detail in Ref.${\,}$\ocite{Restoring}, with
explicit calculations for the optical EPR correlations observed in both Orsay and Innsbruck experiments \ocite{Nature-666}.
Now we have already seen a different aspect of Moldoveanu's difficultly with these basic statistical
concepts in section VI above. This is surprising, because the rules of correlation statistics were discovered by Galton
and Pearson over a century ago \ocite{scores-2}, and today we learn about them in high-school. To be sure, I have used these
rules within the setting of geometric algebra, but Moldoveanu appears to be familiar with the language of geometric algebra,
so that does not quite explain his neglect of these rules. In any event, much of his discussion concerning the
phenomenology of my model is marred by his evident failure to appreciate the distinction between raw scores and
standard scores. This is compounded by his lack of appreciation that quantum correlations
are understood within my framework as purely topological effects, not contextual effects.
By contrast, a complete explanation of how these concepts work within my model and how they
lead to predictions matching exactly those of quantum mechanics can be found in Ref.${\,}$\ocite{Restoring}.

\section{What is Asserted in Moldoveanu's Preprints is Not Necessarily What is True}

Moldoveanu's preprints contain numerous assertions about my framework that are simply not true. It may not be worthwhile
to bring them all out, but let me highlight some examples here to show the extent of his misrepresentation:
\begin{enumerate}
\item It is asserted that I start my disproof with different assumptions from those of Bell.
This is clearly false. It is clear from the first two equations of Ref.${\,}$\ocite{disproof} that both Bell and
I start with the variables ${{\mathscr A}({\bf a},\,{\lambda})=\pm\,1}$ and ${{\mathscr B}({\bf b},\,{\lambda})=\pm\,1}$.
Thus there is no difference between the starting assumptions of my disproof and those of Bell. 
\item It is asserted that my counterexample to Bell's theorem is not a strictly mathematical
disproof. But of course it is. As we saw in section IV above, my counterexample \ocite{disproof} contradicts
the mathematical claim made by Bell. 
\item It is asserted that my interpretation of the EPR argument is ``unusual.'' This is plainly false. I have
faithfully followed the standard interpretation of EPR argument as offered, for example, by Clauser and Shimony
\ocite{Clauser-Shimony-666} and\break by Greenberger {\it et al.}${\,}$\ocite{GHSZ-666} (see especially footnote 10 of the
latter). These are standard references in the subject.
\item It is asserted that I unjustifiably identify ``completeness'' with ``parallelizability'' within the context of EPR
argument. This is an oversimplification of my rather subtle argument relating locality to division
algebras \ocite{What-666}.
\item It is asserted that my counterexample does not satisfy the conditions of remote context independence
and remote outcome independence. But it certainly does, as can be readily seen from the very first two
equations of Ref.${\,}$\ocite{disproof}.
\item It is asserted that my papers ``suffer'' from a convention ambiguity, and that I illegally mix conventions
during computations.${\;}$Nothing can be farther from the truth. There is no such ambiguity or mixing within my model.
\item It is asserted that associating a hidden variable to an abstract computation convention is unphysical. That is
certainly true, but that is not what I am doing. As is clear from the discussions in sections II and III above, orientation
of the physical space, which is the hidden variable ${\lambda}$ in my model, has nothing to do with conventions.
\item It is asserted that I make an extraordinary claim without proof with regard to equation (149) of
Ref.${\,}$\ocite{illusion-666}. This is simply not true. I offer detailed proof of how my argument goes
through, especially in the case of 7-sphere.
\item It is asserted that my model for the EPR correlation does not respect the
detector swapping symmetry. That this is simply false can be seen at once from the definitions 
(\ref{17-Joy}) and (\ref{18-Joy}) used above for the local variables of Alice and Bob. Swapping the detectors
${(-\,I\cdot{{\bf a}}\,)}$ and ${(+\,I\cdot{{\bf b}}\,)}$ of Alice and Bob does not change either the measurement
statistics or correlation. Even changing the sense of one of them only induces a statistically inconsequential sign
change in the correlation. Further discussion on this issue can be found in Refs.${\,}$\ocite{reply} and \ocite{Further-666}. 
\item It is asserted that my model predicts the same correlation regardless of the spin state of the particles in the
EPR-Bohm experiment. That this is not true can be seen at once from the four examples discussed in
Ref.${\,}$\ocite{illusion-666}. In particular, the cosine correlation produced by the rotationally invariant singlet
state corresponds
to the correlation between two equatorial points of the 3-sphere in my model, whereas correlations produced by the
rotationally non-invariant Hardy state correspond to correlations among a set of non-equatorial points of the\break
3-sphere. The GHZ correlations, on the other hand, correspond to the equatorial points of the 7-sphere instead of\break
the 3-sphere. In
general different quantum correlations correspond to different sets of points of the 7-sphere \ocite{illusion-666}.
\item It is asserted that my model for the EPR-Bohm correlation does not satisfy the Malus's law for sequential
spin measurements, and therefore it contradicts the experimental observations as well as the predictions
of quantum mechanics for such measurements. This claim is demonstrably false. An explicit proof of the Malus's law
within my model can be found in section V of Ref.${\,}$\ocite{Further-666}, with further elaborations and detailed
examples of specific spin cases in section III or Ref.${\,}$\ocite{What-666}.
The key ingredient in the proof is again the topology of the parallelized 3-sphere.
\end{enumerate}

\section{Conclusion}

I have shown that much of the criticism of my work in Moldoveanu's preprints stems from incorrect understanding
of my local-realistic framework. I have also shown that, contrary to the claims made in his preprints, my framework
is manifestly non-contextual. In particular, quantum correlations are understood within it as purely topological
effects, not contextual effects. Moreover, I have highlighted a number of conceptual and mathematical errors in
Moldoveanu's\break discussion of my framework. Some of these errors are quite elementary. For example, in a couple of
his arguments Moldoveanu identifies the quantity ${[(+\,I\cdot{{\bf a}}\,)\,(\,+\,{\boldsymbol\mu}\cdot{{\bf a'}}\,)]}$ as a
bivector when it is evidently a multivector (in fact it is a quaternion or a rotor---cf. section IX above). 
This misleads him into developing an erroneous argument against my demonstration of the well known twist within a 3-sphere
\ocite{Restoring}\ocite{Eguchi}. Moreover, much of Moldoveanu's analysis of my work appears to be an attempt to
fit my framework into one preconceived conceptual box or another. The picture\break he thereby ends up creating has therefore
little to do with my framework. Furthermore, some of the concerns raised by Moldoveanu are simply rewording of the concerns previously
raised by others, which I have already addressed elsewhere \ocite{reply}. Notwithstanding these difficulties, I have used
this opportunity to further elucidate at least some aspects\break of my local-realistic framework.
I hope this reassures the reader that my work is perfectly cogent and error free.

\section{Notes added to proof --- Responses to Past Criticisms}

\parskip 3pt

\baselineskip 3pt

Following a suggestion by a reviewer and a journal editor, in this addendum I address some of the criticisms of the\break local-realistic model presented in \ocite{frw}. Previously I have responded to such criticisms in these five preprints: \ocite{reply,Response-2,Response-3,Response-4,Response-5}. Here I make brief comments on them, and address related issues raised online in a personal blog by one of the critics.

\vspace{-0.4cm}

\subsection{Two Critiques by Gill}

\vspace{-0.18cm}

In an unpublished preprint \ocite{Gill-1} Gill has attempted to criticize an earlier version of the local-realistic model I have presented in \ocite{frw}. His critique, however, contains quite a few elementary mathematical and conceptual mistakes. For example, the abstract of the first version of his preprint refers to the quantity ${-{\bf a}\cdot{\bf b}-{\bf a}\times{\bf b}}$ as a ``bivector." And even after my detailed explanations of the difference between a cross product and a wedge product, and the difference between a bivector and a multivector within geometric algebra, all subsequent versions of his preprint continue the mistake of referring to the multivector ${-{\bf a}\cdot{\bf b}-{\bf a}\wedge{\bf b}}$ as a bivector, leading to more serious mistakes later on in the critique. I have systematically corrected these mistakes in my responses \ocite{Response-3} and \ocite{Response-5}.

One of the surprising oversights in Gill's critique is the distinction between the detector bivectors ${{\bf D}({\bf a})}$ and ${{\bf D}({\bf b})}$ and the spin bivectors ${-{\bf L}({\bf s},\,\lambda)}$ and ${+{\bf L}({\bf s},\,\lambda)}$ considered in \ocite{frw}, together with the reciprocal relation between them,
\begin{equation}
{\bf L}({\bf n},\,\lambda)\,=\,\lambda\,{\bf D}({\bf n})\,\,\Longleftrightarrow\,\,{\bf D}({\bf n})\,=\,\lambda\,{\bf L}({\bf n},\,\lambda)\,, \notag
\end{equation}
with ${\lambda}$ being the uncontrollable hidden variable in the sense of Bell \ocite{Bell-1964-666}. In other words, the correct representation of EPR-Bohm experiment and the corresponding spin detection processes defined in Eqs.~(58) and (59) are entirely missing in Gill's portrayal of my model. Consequently, what is described in the preprint \ocite{Gill-1} is {\it not} my model at all.

Moreover, Eq.~(4) of Gill's critique makes another serious mistake regarding the physics underlying the EPR-Bohm experiments. It inserts the equation ${{\mathscr A}({\bf a},\,\lambda){\mathscr B}({\bf b},\,\lambda)=(-\lambda)(+\lambda)=-1}$ for all ${\bf a}$ and ${\bf b}$ even when ${{\bf b}\not={\bf a}}$ by identifying ${{\mathscr A}({\bf a},\,\lambda)}$ with ${-\lambda}$ and ${{\mathscr B}({\bf b},\,\lambda)}$ with ${+\lambda}$, despite the fact that no such equation exists in my model. The insertion of this equation not only violates the conservation of spin angular momentum captured in Eqs.~(69) and (70) of \ocite{frw}, but also confuses the measurement results ${{\mathscr A}=\pm1}$ and ${{\mathscr B}=\pm1}$, which occur at remote stations, with the initial state ${\lambda=\pm1}$, which originate at the central source in the overlap of the backward light cones of Alice and Bob. It is evident from Eqs.~(69) and (70) that ${{\mathscr A}{\mathscr B}=-1}$ for ${{\bf b}\not={\bf a}}$ can occur if and only if the said conservation law is violated \ocite{frw}.

In summary, the critique in \ocite{Gill-1} is a straw-man argument that ignores the fact that my Clifford-algebraic approach to strong correlations is based on a {\it relative} orientation of a quaternionic 3-sphere, taken as Bell's local hidden variable. So much so, that Gill actually replaces one of my central equations with one of his own (thereby introducing a sign error), criticizes his own mistaken equation, and then declares that he has refuted my model. Indeed, in Eq.~(2) of his critique an additional ${\lambda}$ is inserted {\it by hand}, in the middle of that equation, but it does not belong there. I have\break explained this in the paragraph that includes Eq.~(36) in my response \ocite{Response-3}. But this mistake in \ocite{Gill-1} remains uncorrected. 

In a second paper \ocite{Gill-2} Gill criticizes my proposed experiment to test the relevance of Bell's theorem in a macroscopic setting \ocite{IJTP}. Unfortunately, this critique too contains surprisingly elementary mathematical and conceptual mistakes. For example, in the very equation of mine that this critique claims to be criticizing (namely, the standard definition of the bivector subalgebra \ocite{Clifford}), Gill forgets to sum over the bivector-index, arriving at a rather strange conclusion. What is more, the Bell-CHSH correlator is also calculated incorrectly in \ocite{Gill-2}, by summing over spin detections of physically incompatible experiments. I have explained these and further errors in my response \ocite{Response-5} and analysis \ocite{Bell-oversight}. 

\vspace{-0.4cm}

\subsection{Three Critiques by Moldoveanu}

\vspace{-0.18cm}

The two unpublished critiques \ocite{M-1,M-2} of my model are quite similar to the one by Gill discussed above \ocite{Gill-1}. They too are riddled with elementary mathematical and conceptual mistakes \ocite{Response-2}. The argument in these preprints is also a straw-man argument that ignores the physical process underlying the EPR-Bohm experiments as well as the fact that my model is based on a {\it relative} orientation of the quaternionic 3-sphere, taken as a local hidden variable. As I explain below, Moldoveanu too replaces one of my central equations with one of his own (thereby introducing a sign error {\it by hand} in the same manner as Gill), criticizes his own mistaken equation, and then declares that he has refuted my model. I have previously addressed a large number of his claims in my response \ocite{Response-2}. Unfortunately, despite my detailed refutations in \ocite{Response-2}, these same incorrect criticisms have recently appeared in his private blogpost \ocite{MoldyBlog}. These online comments are also at best a series of misunderstandings of the calculations presented in \ocite{frw}. It is, however, instructive to go through these comments one-by-one to understand the mistakes in the blogpost \ocite{MoldyBlog}, as follows. 

\underbar{{\sl Mistake} \#1}: It is claimed in the blog \ocite{MoldyBlog} that my paper has nothing to do with Friedmann-Robertson-Walker spacetime. But this claim is mistaken for several reasons, as explained in the following. 

To begin with, in the context of Bell's theorem the question of local causality is properly addressed only within an adequately relativistic picture of spacetime. See, for example, the lucid discussion by John S. Bell himself in his last paper on the subject \ocite{Bell-1990}. In this paper Bell defines local causality within a given spacetime as follows:
\begin{quote}
A theory will be said to be locally causal if the probabilities attached to values of local beables in a space-time region 1 are unaltered by specification of values of local beables in a space-like separated region 2, when what happens in the backward light cone of 1 is already sufficiently specified, for example by a full specification of all local beables in a space-time region 3 (figure 6.4).
\end{quote}
Moreover, as is well known, a violation of the relativistic local causality can be separated into two conceptually distinct parts: (1) a signalling non-locality incompatible with special relativity, and (2) a no-signalling non-locality compatible with special relativity \ocite{peaceful}. These two conceptually distinct parts are kinematically captured by Bell in his definitions ${{\mathscr A}({\bf a},\,\lambda)}$ and ${{\mathscr B}({\bf b},\,\lambda)}$ of local measurement functions for any given initial state ${\lambda}$ of the system \ocite{Bell-1964-666}\ocite{frw}. This separates relativistic local causality into independence of the parameter ${\bf a}$ from ${\bf b}$ (and vice versa) preserving signalling locality, and independence of the outcome ${\mathscr A}$ from ${\mathscr B}$ (and vice versa) preserving no-signalling locality, in any EPR-Bohm type experiment. This separation allows one to recognize that quantum mechanics preserves {\it parameter independence} (thus remaining compatible with special relativity) but violates {\it outcome independence} (cf. Ref.~\ocite{peaceful}). Thus, despite appearances, relativistic causality is implicit and essential in any discussion involving Bell-type measurement functions.

Now, by definition, the singlet correlations in any EPR-Bohm type experiment are computed among measurement events that occur {\it simultaneously}, at equal times. In practice, this amounts to averaging over ``coincidence counts" of simultaneously  occurring spin detections at spacelike distances, confined to a spacelike hypersurface within spacetime:
\begin{equation}
{\cal E}({\bf a},\,{\bf b})\,=\lim_{\,n\,\gg\,1}\left[\frac{1}{n}\sum_{k\,=\,1}^{n}\,{\mathscr A}({\bf a},\,{\lambda}^k)\;{\mathscr B}({\bf b},\,{\lambda}^k)\right]\equiv\,\frac{\Big[C_{++}({\bf a},\,{\bf b})\,+\,C_{--}({\bf a},\,{\bf b})\,-\,C_{+-}({\bf a},\,{\bf b})\,-\,C_{-+}({\bf a},\,{\bf b})\Big]}{\Big[C_{++}({\bf a},\,{\bf b})\,+\,C_{--}({\bf a},\,{\bf b})\,+\,C_{+-}({\bf a},\,{\bf b})\,+\,C_{-+}({\bf a},\,{\bf b})\Big]},
\end{equation}
where ${C_{+-}({\bf a},\,{\bf b})}$ etc. represent the number of simultaneous occurrences of detections ${+1}$ along ${\bf a}$ and ${-1}$ along ${\bf b}$, etc., with all vectors being spacelike. This may give a false impression that spacetime is irrelevant for the question of local causality in this context and a three-dimensional hypersurface would be sufficient for the analysis of correlations. We must not forget, however, that the measurement outcomes ${{\mathscr A}({\bf a},\,\lambda)}$ and ${{\mathscr B}({\bf b},\,\lambda)}$ depend, not only on the spacelike vectors ${\bf a}$ and ${\bf b}$, but also on the initial state ${\lambda}$ of the singlet system {\it which originates in the overlap of the backward light-cones of Alice and Bob}, as in Fig.~1 of \ocite{frw}. And it is this initial state ${\lambda}$ originating in the overlap that brings about the measurement outcomes ${{\mathscr A}({\bf a},\,\lambda)}$ and ${{\mathscr B}({\bf b},\,\lambda)}$ for any freely chosen spacelike vectors ${\bf a}$ and ${\bf b}$. Thus relativistic considerations are by no means irrelevant for the understanding of local causality in the context of Bell's theorem. 
 
But what is missing from the relativistic considerations by Bell in \ocite{Bell-1990} are the algebraic, geometrical and topological properties of the physical space within which we are confined to perform all our experiments. And that is where the spacelike hypersurface, ${S^3}$, of a Friedmann-Robertson-Walker spacetime enters our analysis in Ref.~\ocite{frw}. As explained in that paper, the geometry of the quaternionic 3-sphere is essential for the derivation of strong correlations, and that geometry is provided by the spacelike hypersurface of one of the three cosmological solutions of Einstein's field equations, as captured in Eq.~(9) of \ocite{frw}. Thus, Friedmann-Robertson-Walker spacetimes are just the right spacetimes for addressing the question of local causality. They are sufficiently Newtonian to host the strong correlations predicted by the singlet state, and sufficiently relativistic to address the question of no-signalling non-locality that appears to be occurring at a spacelike distance in the EPR-Bohm experiments, with the vital condition for local causality being that the initial state ${\lambda}$ must originate in the overlap of the backward light cones of Alice and Bob. Suppose, however, we ignore Eq.~(9) of \ocite{frw} and start the analysis from Eq.~(10) instead. But removing the Friedmann-Robertson-Walker spacetime from the analysis in this manner would make the entire analysis {\it ad hoc}, with no physical justification for ${S^3}$. Thus, the claim that my paper has nothing to do with Friedmann-Robertson-Walker spacetime is quite wrong.

\underbar{{\sl Mistake} \#2}: It is claimed in the blog \ocite{MoldyBlog} that in Eq.~(49) of Ref.~\ocite{frw}, namely, in the standard bivector subalgebra   
\begin{equation}
{\bf L}({\bf a},\,\lambda)\,{\bf L}({\bf b},\,\lambda)\,=\,-\,{\bf a}\cdot{\bf b}\,-\,{\bf L}({\bf a}\times{\bf b},\,\lambda)\,, \notag
\end{equation}
two different algebras are combined into the same equation. The claim is that the bivectors appearing in the above equation are not of the same kind, but a mixture of bivectors corresponding to two different algebraic representations. But this claim is manifestly incorrect. Regardless of ${\lambda}$, all three bivectors ${{\bf L}({\bf a},\,\lambda)}$, ${{\bf L}({\bf b},\,\lambda)}$, and ${{\bf L}({\bf a}\times{\bf b},\,\lambda)}$ in the above equation belong to the {\it same} algebraic representation of the standard bivector subalgebra (48). Thus, contrary to the claim, Eq.~(49) does not describe two different multiplication rules but the same multiplication rule of the standard bivector subalgebra. The mistaken claim stems from a failure to understand what ${\lambda}$ stands for within ${S^3}$. It represents an orientation of the spin bivectors ${{\bf L}({\bf n},\,\lambda)}$ {\it relative} to the detector bivectors ${{\bf D}({\bf n})}$, as defined in Eq.~(60). The meaning of ${\lambda}$ and the relationship between ${{\bf L}({\bf n},\,\lambda)}$ and ${{\bf D}({\bf n})}$ are clearly brought out between Eqs.~(81) and (92). They show that the left-handed subalgebra can be easily transformed into a right-handed subalgebra by reversing the order of the bivectors in their product, as verified also in the numerical simulations with a GAViewer program \ocite{Wonnink,Diether-1,Diether-2}.

\underbar{{\sl Mistake} \#3}: It is claimed in the blog \ocite{MoldyBlog} that matrix representation of the bivector subalgebra using Pauli matrices is equivalent to the usual bivector representation of the subalgebra under consideration. While there is some element of truth in this claim, the matrix representation of bivector subalgebra fails at the very first step, because a product of two Pauli matrices can at most be an identity matrix, not a scalar number. On the other hand, what are observed in the experiments, as results of the interactions between the spins ${{\bf L}({\bf n},\,\lambda)}$ and the detectors ${{\bf D}({\bf n})}$, are pure scalar numbers: ${{\mathscr A}({\bf a},\,{\lambda})=\pm\,1}$ and ${{\mathscr B}({\bf b},\,{\lambda})=\pm\,1}$. Thus a matrix representation is of no use in the present context. 

\underbar{{\sl Mistake} \#4}: It is claimed in the blog \ocite{MoldyBlog} that in Eqs.~(71) to (79) of this paper I am summing over two different representations of the bivector subalgebra in a single sum. But it is quite evident from these equations that what is being averaged over are the measurement results ${{\mathscr A}({\bf a},\,{\lambda})=\pm\,1}$ and ${{\mathscr B}({\bf b},\,{\lambda})=\pm\,1}$, which are limiting scalar points of a quaternionic 3-sphere as defined in the Eqs.~(58) and (59). Consequently, from Eqs.~(71) and (76) we have the following geometrical and statistical identity:
\begin{equation}
\lim_{\,n\,\gg\,1}\left[\frac{1}{n}\sum_{k\,=\,1}^{n}\,
{\mathscr A}({\bf a},\,{\lambda}^k)\;{\mathscr B}({\bf b},\,{\lambda}^k)\right]=\lim_{\,n\,\gg\,1}\left[\frac{1}{n}\sum_{k\,=\,1}^{n}\,{\bf L}({\bf a},\,\lambda^k)\,{\bf L}({\bf b},\,\lambda^k)\,\right].
\end{equation}
Evidently, all bivectors ${{\bf L}({\bf a},\,\lambda)}$ and ${{\bf L}({\bf b},\,\lambda)}$ in this identity belong to the {\it same} algebraic representation of the bivector subalgebra. The mistaken claim results from the previous two mistakes in \ocite{MoldyBlog}. In fact, the steps from (71) to (76) are quite straightforward and have been carefully explained just below Eq.~(79). The steps from (76) to (79) are also straightforward. They follow at once upon using the relation (60). While there is no room for a mistake in these latter three steps, they can be avoided by following Eqs.~(91) to (93) instead, which provide an independent confirmation of the derivation from (71) to (79). Not surprisingly, both calculations give one and the same result (79). What is more, two programmers have independently confirmed the validity of the derivation from (71) to (79) in two event-by-event numerical simulations of the singlet correlations using a GAViewer program based on Geometric Algebra \ocite{Wonnink,Diether-1,Diether-2}. 

This raises a question: Where does the claim of a result different from (79) stem from? It stems from an attempt to insert an additional ${\lambda}$ into the model, {\it by hand}, without a meaningful justification for it. To be sure, the critique tries to justify the additional ${\lambda}$, but without considering either what ${\lambda}$ stands for in the model or the relation (60) between the spin bivectors ${{\bf L}({\bf n},\,\lambda)}$ and the detector bivectors ${{\bf D}({\bf n})}$ it represents. The actual ${\lambda}$ in the model is not a convention but a hidden variable that originates from a central source. The additional ${\lambda}$, on the other hand, is neither originated at the source nor detected by the detectors. It is inserted {\it by hand}. Moreover, contrary to the rationale for inserting an additional ${\lambda}$, there is no such thing as ``orientation independent" or ``orientation-free" objects in Geometric Algebra as claimed in \ocite{MoldyBlog}. An orientation of a vector space such as ${Cl_{3,0}}$ is a {\it relative} concept (cf. the textbook definition of orientation in section 5 of Ref.~\ocite{IJTP}). If ${{\bf B}_1}$ and ${{\bf B}_2}$ are two bivectors related by the orientation ${\lambda}$ as in ${{\bf B}_1=\lambda{\bf B}_2}$, then ${{\bf B}_2=\lambda{\bf B}_1}$ by arithmetic necessity, because ${\lambda^2=1}$. Thus the attempt of inserting an additional ${\lambda}$ into my model based on a narrative of ``orientation independent" versus ``orientation dependent" objects is seriously mistaken. For neither the spin bivectors ${{\bf L}({\bf n},\,\lambda)}$ nor the detector bivectors ${{\bf D}({\bf n})}$ actually depend on ${\lambda}$. They are only {\it related} by it. Thus the inclusion of additional ${\lambda}$ is a pure fiction that has nothing to do with the actual ${S^3}$ model presented in \ocite{frw}.

\underbar{{\sl Mistake} \#5}: It is claimed in the blog \ocite{MoldyBlog} that ``correlations must be computed using actual experimental results" and ``must not be made in a hypothetical space of `beables'." 

Yes, correlations must be computed using actual experimental results of ${+1}$ and ${-1}$, {\it but only to the extent that quantum mechanics is able to predict such actual measurement results}. After all, any local-realistic theory is obliged to reproduce only that which quantum mechanics is able to predict statistically and experimentalists are able to observe experimentally \ocite{Clauser-Shimony-666}. So, with that important correction to the claim, the correlations are indeed computed in the paper using actual experimental results of ${+1}$ and ${-1}$. Such actual experimental results are explicitly specified by the limiting scalar points ${{\mathscr A}({\bf a}\,,\,\lambda)=\pm1}$ and ${{\mathscr B}({\bf b}\,,\,\lambda)\pm1}$ of a quaternionic 3-sphere, which models the physical space in which we are confined to perform all our experiments. They correspond exactly to the measurement results considered by Bell in his paper (cf. Eq.~(1) of Ref.~\cite{Bell-1964-666} and Eqs.~(58) and (59) of Ref.~\ocite{frw}). These ${+1}$ or ${-1}$ results are then averaged over in Eq.~(71), which is {\it the} standard way of computing the correlations in the experimental context of Bell's theorem. No ``hypothetical space of `beable'" is involved in this paper, or anywhere else in my work.

\underbar{{\sl Mistake} \#6}: It is claimed in the blog \ocite{MoldyBlog} that ``James Weatherall found a mathematically valid example very similar with [my] proposal but one which does not use quaternions/Clifford algebras." But this claim is not correct, as explained in the following paragraph.

\vspace{-0.4cm}

\subsection{A Critique by Weatherall}

\vspace{-0.18cm}

Despite its claim, Weatherall's critique \ocite{Weatherall} is not a critique of my local-realistic model at all but merely an exposition of the standard Bell's theorem \ocite{Bell-1964-666,Bell-1990}. The critique begins by giving the wrong impression that the author is about to present and criticize my quaternionic 3-sphere model for the strong correlations. But, in fact, it does no such thing. The critique immediately switches to a different, non-Clifford-algebraic model based on an ordinary 2-sphere\footnote{It is well known in algebraic topology that, unlike on a 3-sphere, hair on a 2-sphere cannot be combed without creating a cowlick.\label{cow}} instead of a quaternionic 3-sphere, and shows that his unphysical model does not reproduce the strong correlations. But it is quite well known for more than fifty years that any na\"ive attempt which ignores the correct algebra, geometry, and topology of the compactified physical space (${S^3}$) cannot reproduce the strong correlations between the measurement results such as ${{\mathscr A}=\pm1}$ and ${{\mathscr B}=\pm1}$. What is more, in an unnumbered equation Weatherall makes the same mistake regarding the conservation law underlying any EPR-Bohm experiment that Gill has made in his critique discussed above \ocite{Gill-1}. He inserts the product ${{\mathscr A}({\bf a},\,\lambda){\mathscr B}({\bf b},\,\lambda)=(-\lambda)(+\lambda)=-1}$ for all ${\bf a}$ and ${\bf b}$ even when ${{\bf b}\not={\bf a}}$, despite the fact that no such equation appears in my model. This equation not only violates the conservation of spin angular momentum captured in Eqs.~(69) and (70) of \ocite{frw}, but also confuses the measurement results ${{\mathscr A}=\pm1}$ and ${{\mathscr B}=\pm1}$, which are observed at remote detectors, with the initial state ${\lambda=\pm1}$, which originates at the central source. Moreover, it is evident from Eqs.~(69) and (70) of \ocite{frw} that ${{\mathscr A}{\mathscr B}=-1}$ for ${{\bf b}\not={\bf a}}$ can occur if and only if the conservation of spin angular momentum is violated. Notwithstanding these facts, Weatherall argues that, since his non-Clifford-algebraic model based on a non-combable\footnotemark[\getrefnumber{cow}] 2-sphere fails, my Clifford-algebraic model of the correlations must also fail, without even mentioning a 3-sphere or a quaternion, and without pointing to any mistake in my explicit and constructive model. In my response in \ocite{Response-4} I have explained the above shortcomings of his critique in greater detail. More recently, I have also pointed out an oversight in a Bell-type unphysical argument on which Weatherall's critique depends \ocite{Bell-oversight}.

\vspace{-0.4cm}

\subsection{Summary of the Failures of Criticisms}

\vspace{-0.18cm}

A common flaw in all of the critiques discussed above is the omission of considering the correct physical process in any EPR-Bohm type experiment. What is involved in such experiments are two prearranged detectors of Alice and Bob with the same orientation, stationed at remote locations, which I have represented as bivectors ${{\bf D}({\bf a})}$ and ${{\bf D}({\bf b})}$, and two randomly oriented spins generated from a central source, which I have represented as spin bivectors ${-{\bf L}({\bf s},\,\lambda)}$ and ${+{\bf L}({\bf s},\,\lambda)}$ (cf. Fig.~2 in Ref.~\ocite{frw}). What is observed are then simultaneous interactions between the detectors ${{\bf D}({\bf a})}$ and ${{\bf D}({\bf b})}$ and the spins ${-{\bf L}({\bf s},\,\lambda)}$ and ${+{\bf L}({\bf s},\,\lambda)}$, at spacelike distances. The Clifford-algebraic calculations are then relatively straightforward \ocite{frw}. It is the failure to take into account these basic features of the EPR-Bohm type experiments within an appropriate Clifford-algebraic setting that has lead the critics to their mistaken conclusions. 

Now, as Moldoveanu claims on his blog, {\it Annals of Physics}, did remove my paper from its website after a month of its publication, within minutes of receiving a complaining email from Richard D. Gill. However, the journal did not notify me about the removal for over two months, and, despite my repeated requests, {\it has not provided any evidence of a mistake in the paper -- even privately}. On the web page of the paper the journal states that ``...the results [of my paper]\break are in obvious conflict with a proven scientific fact, {\it i.e.}, violation of local realism that has been demonstrated not only theoretically but experimentally in recent experiments. On this basis, the Editors decided to withdraw the paper."

This is quite an extraordinary statement, not the least because my paper went through seven months and two rounds of rigorous peer review, but no editor or reviewer were able to detect the alleged ``obvious conflict." Moreover, since the framework for the EPR-Bohm correlations presented in the paper reproduces all quantum mechanical predictions and experimental results for the singlet state {\it exactly}, the journal's claim of ``obvious conflict" is manifestly wrong.

\vspace{-0.2cm}

\acknowledgments

I wish to thank Fred Diether and Lucien Hardy for their comments on an earlier version of the manuscript. I also wish to thank the Foundational Questions Institute (FQXi) for supporting this work through a Mini-Grant.

\vspace{-0.2cm}

\renewcommand{\bibnumfmt}[1]{\textrm{[#1]}}

\end{document}